\documentclass[10pt,journal,compsoc]{IEEEtran}
\ifCLASSOPTIONcompsoc
  \usepackage[nocompress]{cite}
\else
  \usepackage{cite}
\fi
\ifCLASSINFOpdf
   \usepackage[pdftex]{graphicx}
\graphicspath{{figures/}{pictures/}{images/}{./}} 
\else
   \usepackage[dvips]{graphicx}
\fi

\graphicspath{{figures/}{pictures/}{images/}{./}} 

\newcommand{\squeezeup}{\vspace{-4mm}}

\usepackage{color}

\begin{document}
%
\title{WeSeer: Visual Analysis for Better Information Cascade Prediction of WeChat Articles}
%
%
%
%

\author{
	    Quan Li,
	    Ziming Wu,
	    Lingling Yi,
	    Kristanto Sean N,
	    Huamin Qu,~\IEEEmembership{Member,~IEEE}
	    and Xiaojuan Ma
\IEEEcompsocitemizethanks{
	\IEEEcompsocthanksitem
	Quan Li, Ziming Wu, Kristanto Sean N, Huamin Qu, and Xiaojuan Ma are with the Hong Kong University of Science and Technology.
	E-mail: \{qliba, zwual, huamin, mxj\}@cse.ust.hk; kristantosean@gmail.com.
    \IEEEcompsocthanksitem
    Lingling Yi is with WeChat, Tencent.
    E-mail: chrisyi@tencent.com.
}
}

%
%

\markboth{IEEE TRANSACTIONS ON VISUALIZATION AND COMPUTER GRAPHICS}%
{Shell \MakeLowercase{\textit{et al.}}: Bare Demo of IEEEtran.cls for Computer Society Journals}
%



\IEEEtitleabstractindextext{%
\begin{abstract}
Social media, such as Facebook and WeChat, empowers millions of users to create, consume, and disseminate online information on an unprecedented scale. The abundant information on social media intensifies the competition of WeChat Public Official Articles (i.e., posts) for gaining user attention due to the zero-sum nature of attention. Therefore, only a small portion of information tends to become extremely popular while the rest remains unnoticed or quickly disappears. Such a typical ``long-tail'' phenomenon is very common in social media. Thus, recent years have witnessed a growing interest in predicting the future trend in the popularity of social media posts and understanding the factors that influence the popularity of the posts. Nevertheless, existing predictive models either rely on cumbersome feature engineering or sophisticated parameter tuning, which are difficult to understand and improve. In this paper, we study and enhance a point process-based model by incorporating visual reasoning to support communication between the users and the predictive model for a better prediction result. The proposed system supports users to uncover the working mechanism behind the model and improve the prediction accuracy accordingly based on the insights gained. We use realistic WeChat articles to demonstrate the effectiveness of the system and verify the improved model on a large scale of WeChat articles. We also elicit and summarize the feedback from WeChat domain experts.
\end{abstract}

}

\maketitle

\IEEEdisplaynontitleabstractindextext

%
\IEEEpeerreviewmaketitle

\IEEEraisesectionheading{\section{Introduction}\label{sec:introduction}}

%
%
%
%

\IEEEPARstart{W}eChat\footnote{http://www.wechat.com/en/}, which is a new type of social networking service, has already become ubiquitous in the daily mobile communication of the Chinese. Reports\footnote{http://www.businessinsider.com/wechat-has-hit-1-billion-monthly-active-users- 2018-3} indicate that by the end of March 2018, WeChat had approximately 1,040 million monthly active users. In addition to its typical functions, such as direct messaging, WeChat allows individuals and organizations to register official accounts for publicizing articles (i.e., posts). Reading and sharing articles have become important activities of WeChat users.

\par Predicting the ultimate popularity of WeChat articles is important. First, the abundant information on WeChat intensifies the competition of articles for gaining user attention due to the zero-sum nature of attention. Therefore, only a small portion of information becomes extremely popular, whereas the rest remains unnoticed or rapidly disappears. For example, statistical results show that, although 1.5 million articles are generated each day, only 0.07\% of the total articles are shared over 10,000 times. Popularity prediction allows WeChat to rank articles for recommendation purpose, discover potential trending articles, and improve its social recommendation applications. Second, articles with anomalous behaviors and purposes can be potential threats to the online communication system. WeChat needs to trace the spread of articles, make decisions, and act at an early stage before a large popularity propagation to control the spread of rumors. Third, predicting the future trend of articles allows understanding of the factors that influence the popularity of the articles, gaining insights into collective behaviors, and carrying out efficient advertising campaigns.

\par A team of experts from WeChat conventionally applies feature-based models~\cite{qiu2016lifecycle} to estimate the probability of large propagation for each article and mark articles with a predictive final propagation size greater than a predetermined threshold. However, shortcomings are inherent in the approaches used at present. \textbf{(1) Demanding Feature Engineering.} For the feature-based models, an exhaustive set of potentially effective features and even their weights need to be extracted and determined to predict the future growth of information propagation. However, feature engineering can be costly and cumbersome because it consumes a large number of articles to capture the necessary features and construct the training/testing set. Moreover, given that numerous and diverse accounts and articles constantly appear, new features may emerge, and the weights of existing ones may change. \textbf{(2) Binary Classification.} Most feature-based models consider propagation prediction as a classification problem, in which a predefined threshold is required to define the so-called ``viral'' articles. Therefore, they only provide a one-time finding and are thus coarse-grained for further analysis. \textbf{(3) Lack of Sensitivity.} The predetermined threshold is based on a selected time window of the input propagation historical data. This empirical choice is questionable because not all final ``viral'' articles exhibit ``viral'' properties in the fixed time window, and not all predicted ``viral'' articles are truly ``viral'' after a certain observation period. \textbf{(4) Skepticism about Performance.} Most feature-based models involve demanding diagnosticity that causes difficulty in understanding the model mechanism, which may affect prediction results; consequently, improving prediction accuracy is also difficult. For example, when the prediction and the truth do not match, the problem is difficult to identify, ``\textit{it is a black box to analysts who don't understand the constraints of the inside mechanisms and how to employ it appropriately}''~\cite{Dove2017UX}. On the contrary, point process-based models~\cite{matsubara2012rise,mohler2011self,SEISMIC:2015:KDD} directly model an individual's behavior and the formation of an information cascade in a network and then aggregate the effects of the individual to make a prediction; thus, they are more light weighted, and their principle is bottom-up. Unlike the feature-based models that merely use features in classifiers to conduct a one-time finding of future growth, the point process-based models can reflect a consistently updating status of propagation trend in real-time online settings and are therefore more fine-grained for detailed propagation analysis.

\par Although point process-based models improve the diagnosticity better than feature-based models, more efforts are directed to parameter tunings of the inner mechanism to obtain a better prediction result~\cite{gao2015modeling,SEISMIC:2015:KDD}. Visualization is recently applied to understand the involved information propagation, diagnose a predictive model mechanism, and provide visual evidence to support or refute conclusions~\cite{ye:2016:visual,cao:2012:whisper,marcus:2011:twitinfo,ho:2011:modeling}, which are theoretically and practically important for experts. However, designing an effective visualization for predictive models in WeChat article scenario is technically and empirically challenging. First, the large amount of sharing records of articles are multidimensional, thereby presenting difficulty in organizing and visualizing them. Second, unfolding the inner mechanism requires different levels of efforts because it may involve extracting different sets of parameters. Even experienced technical experts need to spend considerable time on the inevitable trial-and-error process to indicate appropriate parameters for those growing sorts of WeChat articles, such as breaking news and entertainment, for everyone, in contrast to science that exhibits various manifestations of propagation patterns. Third, few empirical studies have been conducted on the extent to which the gained visual insights can be leveraged to reshape the improvement of prediction accuracy and further enhance the understanding of the reasons behind different prediction results.

\par In this study, we explore how visualization assists in evolving the domain experts' ``conversation''~\cite{ozenc2010support} with a point process-based model. We first review their conventional approaches to propagation prediction, abstract the needs, and find that a point process-based model, SEISMIC~\cite{SEISMIC:2015:KDD}, is preferable. In consideration of the affordance and inadequacy of the model, we propose a novel visual reasoning approach to help refine the parameters that are conventionally trained automatically in a unified setting across articles and derive factors that may lead to varied prediction results of different articles. The visual reasoning approach summarizes the learned rules by moving the snapshot on the time window of the article's historical data to predict future growth. Our visual approach can lead to a better prediction result than the original SEISMIC model. Representative real-world case studies, expert interviews, and quantitative experiments are utilized to evaluate our approach. We summarize the major contributions as follows:
\begin{itemize}
\item We study and enhance a point process-based model for an improved prediction result.
\item We design a visual reasoning system to support communication between the experts and the point process-based model and derive the reasons for the different prediction results of WeChat articles.
\item We verify the efficacy of our proposed method by case studies and also by conducting a quantitive study in comparison to the original model on a large collection of WeChat articles to support the explainability of the generated results.
\end{itemize}

\section{Related Work}
\par The literature that explores several aspects that overlap with those discussed in this work can be sorted into three categories, namely, information popularity prediction, visual analytics of information propagation, and predictive visual analytics on social media.

\subsection{Information Popularity Prediction}
\par Researchers have deciphered the secret of information popularity by analyzing the information diffusion characteristics on some famous microblogging platforms, such as Facebook, Twitter, and Weibo~\cite{cha:2009:measurement,bao:2013:cumulative,bao:2013:popularity,gao:2014:effective,li:2016:exploring}. These works have set a theoretical foundation and stimulated the blooming of research thereafter. With the increase in social media users, the prediction and analysis of the information popularity of web contents have gathered a large number of researchers, and many prediction models for analyzing different sources of web contents have been proposed in recent years~\cite{tatar:2014:survey,gao:2015:modeling,luo:2012:predicting,ding:2017:predicting,bao:2015:modeling}.

\par One type of the pioneering popularity prediction methods is based on features, which extracts an exhaustive list of potentially relevant features. For example, Bandari et al. predicted the popularity of web contents prior to their release by utilizing features derived from article properties as predictors~\cite{bandari:2012:pulse}. Can et al. incorporated image-, content-, and structure-based features for prediction tasks~\cite{can:2013:predicting}. Cheng et al. addressed cascade prediction problems by studying the features of post content, network structure, and temporal evolution~\cite{cheng2014can} and leveraged different learning algorithms, such as regression models~\cite{cheng2014can}, content-based models~\cite{naveed2011bad}, and regression trees~\cite{bakshy2011everyone}. These approaches encounter laborious feature engineering and extensive training. In other words, feature quality affects model performance~\cite{bandari:2012:pulse}.

\par Another category of approach is the point process-based model, which directly studies the information cascade formation in a social network~\cite{matsubara2012rise,mohler2011self,SEISMIC:2015:KDD}. Most of these models are developed to infer the social network structure over which the cascades propagate~\cite{rodriguez2014uncovering,gomez2013structure}. In this study, our goal is to predict the final cascade size. We improve a well-established point process-based model, SEISMIC, in WeChat scenario. Unlike the original model, our approach can predict the final popularity in a more flexible and reliable manner for various types of WeChat articles.

\subsection{Visual Analytics of Information Propagation}
\par The merits of visual analytics, such as providing insight and understanding of information propagation, have been studied for years~\cite{keim:2015:bridging}. Li et al. investigated how information propagation in a specific microblogging platform evolves to identify relevant patterns and understand the dynamic attributes of information propagation and the underlying sociological motivations~\cite{li2013visual}. Chen et al. proposed \textit{D-map} to explore the diffusion of information on social communities in Sina Weibo~\cite{chen2016d}. Lu et al. presented a visual analytic framework for event cueing using media data~\cite{lu2016exploring}. Zhao et al. presented \textit{FluxFlow} to study the spread of anomalous information in social media~\cite{zhao2014fluxflow}. They leveraged machine learning algorithms to detect anomalies and offered novel visual designs to depict the detected threads for a deep analysis. Wu et al. introduced \textit{OpinionFlow} to enable analysts to detect opinion propagation patterns~\cite{wu2014opinionflow}. In short, special designs, such as map-like~\cite{chen2016d} and river-like techniques~\cite{zhao2014fluxflow,wu2014opinionflow} are used to describe the characteristics of propagation properties and the involved messages and users~\cite{chen2017social}.
\par However, these visual analytic systems mainly target information diffusion and visualize the output of the underlying models. In this study, target users are involved in exploring the predictive model, and we mainly focus on the predictive analysis of cascades rather than the information propagation analysis.

\subsection{Predictive Visual Analytics on Social Media}
\par Although researchers have contributed to the predictive analysis by proposing different models or adopting effective features, a gap remains toward a reliable model. How to distinguish and make choices among them and to understand the mechanisms remain open problems~\cite{el:2014:predictive,gleicher:2014:position}.

\par With the understanding of existing behaviors, visual analytic systems help predict future information trends based on an underlying trained model, derive insights by using visualization techniques, or visually verify prediction results~\cite{chen2017social}. For example, M{\"u}hlbacher et al. provided a survey of frequently used algorithms and discussed possibilities to enable user involvement in ongoing computations~\cite{muhlbacher2014opening}. Yeon et al. combined the contextual similar cases in the past to predict and visualize event evolution patterns~\cite{yeon2015predictive}. Their approach combines with a prediction model and visualization. Other visual analytic approaches rely on either the temporal relevant past events~\cite{maciejewski2011forecasting,hao:2011:visual,bosch2013scatterblogs2} or spatially correlated activities~\cite{malik2014proactive}. Few works have addressed predictive analysis in the context of visual analytics~\cite{Lu2017The}. Integrating interaction and visualization tightly in predictive analysis remains challenging~\cite{chen2017social}.

\par A partition-based framework for building and validating regression models is proposed to address such limitations as the selection of input variables, i.e., feature subset selection~\cite{muhlbacher2013partition}. Relationship structures are qualitatively visualized, and their relevance to feature ranking is quantified. Similar to this approach, we also conducted a ``step-wise'' identification of a regression model. However, in this study, we do not target feature/parameter selection problems but visually reason the value ranges of the studied parameters in the predictive model for different WeChat articles. Thus, we can flexibly obtain a reliable prediction result toward different types of articles.

\section{Background and Preliminary Study}
\subsection{Propagation Process of WeChat Articles}
\par
\begin{figure}[h]
    \centering
    \includegraphics[width=\linewidth]{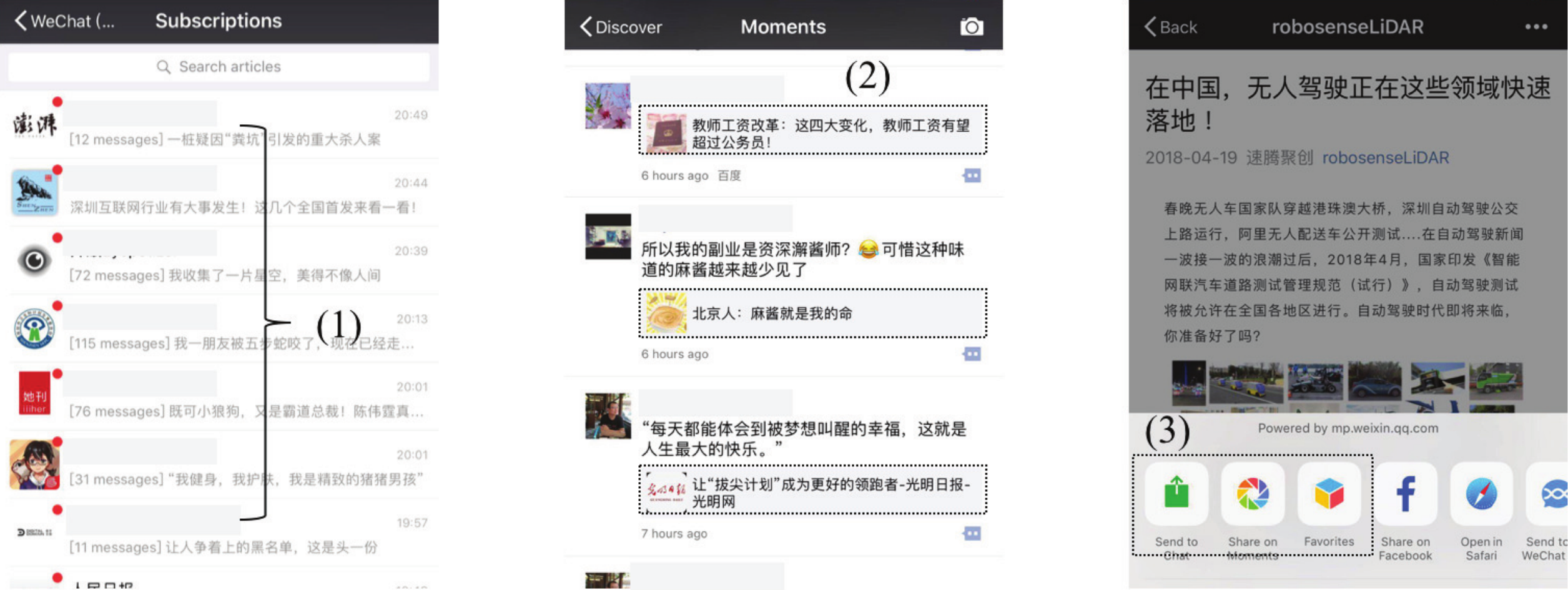} 
        \squeezeup
    \caption{(1) A list of subscription accounts publish articles like a daily news feed; (2) WeChat users share the articles on their Moments; (3) WeChat users can choose to share the article via multiple channels, such as private chatting, group chatting, Moments or adding to favorites.}
    \label{fig:wechat}
    \squeezeup
\end{figure}

 WeChat Official Account Platform serves as the main source of articles for publicizing, reading, and sharing articles. Subscription accounts are often used similarly to daily news feeds because they can push one or several new update(s) to their followers every day. The update(s) can contain a single article or multiple articles bundled together. Users may subscribe to as many accounts as they like. All subscription accounts are placed together in a subscription account folder on the timeline of users. The information cascade of articles in WeChat often follows the ensuring process, i.e., an account publishes an article (Fig.~\ref{fig:wechat} (a)). The followers who read this article may opt to share it, their respective sets of followers will be exposed to this article, and the cascade continues. In WeChat platform, apart from sharing articles on users' Moments (similar to the Facebook timeline) (Fig.~\ref{fig:wechat} (b)), articles can also be shared to a specified friend or a group of friends via direct messaging (Fig.~\ref{fig:wechat} (c)). All the WeChat articles we studied are ``posts''; hence, we refer to the terms of ``posts'' and ``articles'' interchangeably.

\subsection{Experts and their Conventional Practice}
\par We collaborate with three experts from WeChat, including two data scientists (E.1-2) and one business manager (E.3), for three months. They have provided the rule of thumb to define the ``final'' propagation of a certain WeChat article. According to their statistics, 95\% of WeChat articles cannot propagate beyond three days. Therefore, suggested by the experts, we set one week from an article's publication as its final propagation size.
\par E.1-2 are mainly responsible for the information cascade research on WeChat articles. They examine whether the current propagation of a particular article meets the criteria to form a viral propagation. Considering the requirements of generalization and running time, they need to make a tradeoff between selecting features and preserving or improving the discriminative capability of the classifier. They focus on the following three categories of features: \textbf{(1) Article Information.} This category contains the basic information of the WeChat article, including the article length, the posted time, and the number of the subscribers of the article's publisher. \textbf{(2) Sharing Activities.} This category mainly includes the percentage of the ``first-level nodes'' (nodes that directly reshare articles from the root node) within the first $X$ shares, the total number of friends of the first $X$ shares, the number of reads of the articles when they reach the $X^{th}$ share, the number of users who share by direct messaging, and the number of users who share by posting on Moments. \textbf{(3) Sharer Information.} ``Sharer'' is defined as the one who shares the article. The aggregation of the sharers, such as the age distribution, the gender, and the geographical distribution, reflects the aggregation degree of the sharers. Given the training set of first $X$ shares of $M$ articles, the prediction task is to determine whether each article will achieve the median final cascade size of all $M$ articles. They then apply classification models, such as SVM to select the possible articles that may propagate virally. Eventually, E.3 combines the analysis and observation of E.1-2 to determine and promote the articles that should be monitored.
\par Although feature engineering can improve the performance of classifiers, the team encounters the following difficulties when engaging it to the practical scenarios: \textbf{(1) Sensitivity.} Feature-based models typically treat cascade prediction as a classification problem, which defines ``thresholds'' that differentiate ``viral'' articles. However, these ``thresholds'' are attained on the basis of a fixed time window of the historical data of input propagation from various articles, and they may be insufficiently sensitive to the propagation prediction task for a certain type of article. In other words, the current setting cannot capture the temporal information of the sharing process, which may vary across different articles. \textbf{(2) Flexibility.} The collaboration with WeChat enables us to learn that three main propagation patterns exist in typical article cascades: \textit{(a) Immediate Outbreak.} Such articles as breaking news or high-quality articles may experience a sudden increase. \textit{(b) Rise and Recession.} Articles that mainly propagate through Moments and Group Chatting may have relatively lengthy lifecycles. \textit{(c) Wave-like Persistent Propagation.} Such articles as superstition-related topics may propagate in this manner. Sweeping all types of articles in one unified setting will not allow adaptive thresholds. \textbf{(3) Diagnosticity.} The mechanism of feature-based models is usually difficult to understand. Consequently, prediction accuracy is also difficult to improve, and the factors that may affect prediction results are difficult to determine.
\par We determine that a ``simple and principle bottom-up''~\cite{SEISMIC:2015:KDD} model of cascade behaviors is preferable to resolve the above-mentioned issues. A well-established point process-based model, SEISMIC, has recently been proposed to predict the final size of shares for a given tweet~\cite{SEISMIC:2015:KDD}. This model enables infectiousness to vary over time, which indicates that infectiousness can decrease as the tweet content becomes stale. Moreover, whether the current cascade is in a ``supercritical'' or ``subcritical'' state can be identified. A cascade is exploding if it is in the ``supercritical'' state, and its final size cannot be predicted. By contrast, the final size can be predicted by using the Galton-Watson tree~\cite{durrett:2010:probability} if the cascade is in the ``subcritical'' state. We approach the issue of the cascade prediction of WeChat articles as a regression task and opt for the SEISMIC model on the basis of the following merits: \textbf{(1) Generality.} A complete social network structure is usually difficult to obtain and handle, especially for WeChat that involves billions of nodes and edges. SEISMIC only assumes minimal knowledge of the network, namely, limited history of shares and degrees (number of friends) of the sharing nodes; thus, costly feature engineering is not required. \textbf{(2) Scalability.} SEISMIC requires computational time only linear in the number of observed shared of each article; therefore, this model can easily provide immediate feedback on each individual article prediction. \textbf{(3) Accuracy.} As shown in~\cite{SEISMIC:2015:KDD}, this model is more robust and accurate than the time-series linear regression and the two other point process-based methods, i.e., dynamic and reinforced poisson models, when applied to tweet data. In the following subsections, we briefly introduce the SEISMIC model to understand its affordance and inadequacy when we apply it to predict the cascade size for WeChat articles.

\subsection{SEISMIC Model}
\par The goal of SEISMIC is to predict the final propagation size $R_{\infty}$ of an article. Important variables of SEISMIC are $R_t$, i.e., the total number of shares of a given article up to time $t$, and $\lambda_t$, i.e., the current spread rate (intensity) of the cascade. SEISMIC introduces two important components, namely, infectiousness $p_t$ and human reaction time, to determine the spread rate $\lambda_t$ of the cascade. \textbf{(1) Infectiousness.} SEISMIC assumes that each article has a ``time-dependent, intrinsic''~\cite{SEISMIC:2015:KDD} infectiousness parameter $p_t$, which represents how likely an article will be shared at time $t$. \textbf{(2) Human Reaction Time.} Human reaction time models the delay between an article being posted to a user (e.g., appearing in the user's feed) and the time when that user shares that article. This factor is modeled as a ``probability density'' $\phi(s)$ (i.e., \textit{memory kernel}~\cite{crane:2008:robust}), where $s$ is the time difference between the current time and the share's initial post time. With the two components, the article's intensity $\lambda_t$ that represents the probability of obtaining a share at time $t$ can be calculated as
\begin{equation}
\lambda_t=p_t \cdot \sum_{t_i \leq t}n_i\phi(t-t_i),
\end{equation} where $n_i$ is the degree of node $i$ (i.e., users that can see share \textit{i}). Therefore, the likelihood that share \textit{i} occurs at $t_i$ is given by
\begin{equation}
P(t_i | t_1, ..., t_{i-1}) = \lambda_{t_i} \cdot exp(-\int_{t_{i-1}}^{t_i}\lambda_{s}ds).
\end{equation}
The total likelihood is:
\begin{equation}
P(R_t=r,t_1, ..., t_r) = \prod_{i=1}^{R_t}\lambda_{t_i} \cdot exp(-\int_{t_0}^t\lambda_sds),
\end{equation} where $R_t$ is the current number of shares. The infectiousness $p_t$ can be estimated by maximizing the likelihood as
\begin{equation}
p_t=\frac{R_t}{\sum_{i=1}^{R_t}n_i\int_{t_i}^t\phi(s-t_i)ds} .
\end{equation}
The denominator can be interpreted as the ``effective'' number of exposed users, denoted as $N_t^e$. The mean degree of the network is represented as $n_*=(1/R_{\infty})\sum_{i=0}^{R_{\infty}}n_i$, and SEISMIC assumes that node degree \{$n_i$\} is independent and identically distributed with mean degree $n_*$. ${1}/{n_*}$ is the infectiousness as $t \to \infty$.

\begin{figure}[h]
    \centering
    \includegraphics[width=\linewidth]{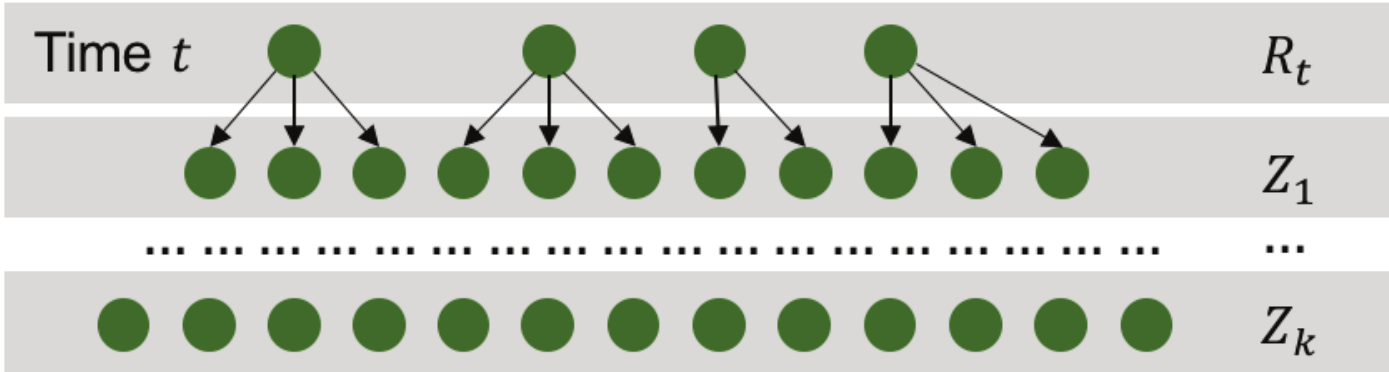} 
    \squeezeup
        \squeezeup
    \caption{In SEISMIC, the cascade up to time \textit{t} is observed and we predict how the cascade tree will grow in the future. $Z_k$ denotes the number of shares caused by the $k^{th}$ generation descendants.}
    \label{fig:diffusion}
\end{figure}

\par After infectiousness $p_t$ is calculated, the final propagation size can be predicted. Let $Z_k$ be the number of shares in the $k^{th}$ generation descendants. The final sharing count is simply $R_t+\sum_{i=1}^\infty Z_k$ (see Fig~\ref{fig:diffusion}). The final propagation size according to probability theory is
\begin{equation}
R_\infty = R_t + \frac
{p_t(N_t - N_t^e)}{1-p_tn_*}.
\end{equation}
\par SEISMIC adopts two evaluation metrics to measure the accuracy of the prediction model. \textbf{(1) Absolute Percentage Error (APE).} Given an article \textit{w} and a prediction time \textit{t}, this metric is defined as
\begin{equation}
APE(w,t)=\frac{|Pre_{\infty}(w,t)-R_{\infty}(w)|}{R_{\infty}(w)}.
\end{equation} \textbf{(2) Breakout Article Coverage.} Large cascades are usually important to real-world business; hence, predicting whether the final cascade size of an article is within the top $M$ of all articles will be useful. In this study, we collect a ground-truth list of top 100 articles ranked by their final size. We also produce a predicted top 100 list based on our methods. We then evaluate the methods by quantifying the degree of overlap between the predicted top 100 list and the ground-truth top 100 list. We adopt the two evaluation metrics to conduct performance comparison between SEISMIC and our approach (Subsection 5.3).

\begin{figure*}[h]
    \centering
    \includegraphics[width=\linewidth]{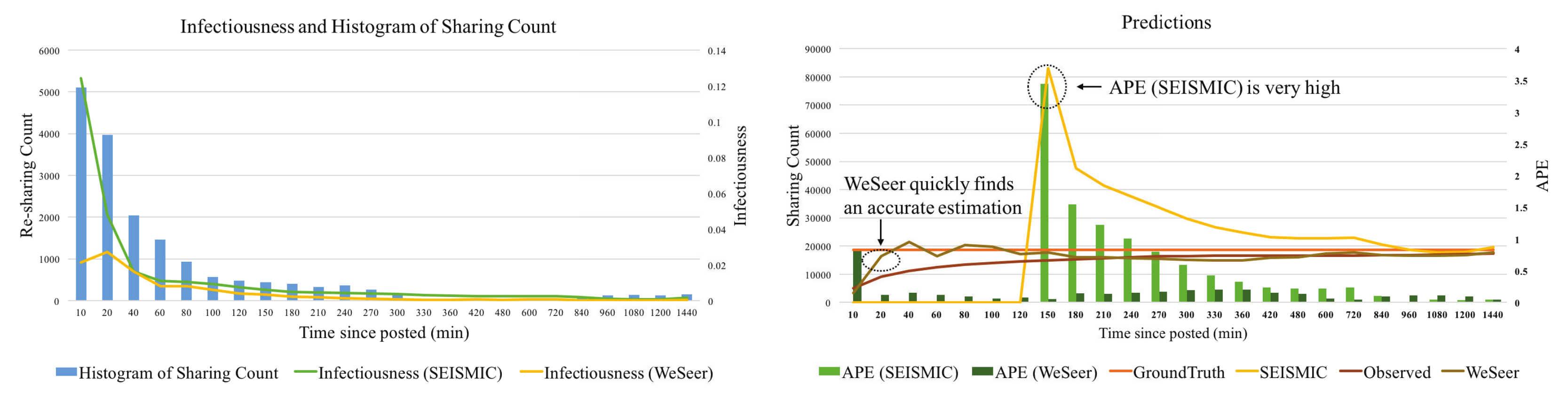} 
        \squeezeup
                \squeezeup
    \caption{One-day sharing activity of a WeChat article which is about the release of the name list of ``double first-class'' university in China. The left figure shows that the propagation first increases dramatically and then quickly decreases (see the histogram of sharing count). The article infectiousnesses of SEISMIC and WeSeer (introduced later) are estimated and compared (see infectiousness curves). The right figure shows that the APEs of SEISMIC and WeSeer at each timestamp, the predictions of final propagation size of SEISMIC and WeSeer, and the true final size (``GroundTruth''). ``Observed'' curve plots the cumulative number of observed shares by a given time.}
    \label{fig:seismic}
    \squeezeup
\end{figure*}

\subsection{Preliminary Study}
\par We conduct a preliminary study to compare the performance of a feature-based model, i.e., SVM, and SEISMIC on predicting the propagation popularity of WeChat articles. We measure the median accuracy explained in Subsection 3.2, i.e., for a certain threshold $X$, we calculate the prediction results, classify them to be either below or above the median value, and compare them with the final actual results.

\par We randomly collect 7509 articles that were published during the first week of March 2016 and obtain their sharing histories until the first week of April 2016. We have three main tables, namely, sharing, reading, and user. The sharing table consists of the sharing activities of the articles, including article id, user id, sharing timestamp, and article post time. The reading table includes article-reading activity. This table has similar columns to those of the sharing table. The user table consists of the demographic information of WeChat users who are involved in the sharing and reading activities. We also have information about article publishers, such as the number of subscribers.

\par We run SVM for thresholds $X$ = 200, 300, 500, 800, 1000. $X$ is the number of shares we have observed. We use cross-validation to measure the accuracy in predicting whether an article will reach the median cascade size. We then implement SEISMIC using the exact same parameters, including $\phi(s)$, $p_t$, $n_*$, and apply it to our dataset. The results are shown in Table~\ref{tab:svm}. The median accuracy of SEISMIC is mostly higher than that of SVM. Without feature engineering, SEISMIC can at least achieve a good performance in cascade size prediction as the feature-based SVM considering that we only leverage the same parameters in the original SEISMIC.

\begin{table}[h]
\centering
\squeezeup
\caption{Median accuracy comparison of SVM and SEISMIC for threshold $X$ = 200, 300, 500, 800, and 1000, respectively.}
\begin{tabular}{|l|cccc|}
\small X & \small \# of articles & \small CV of SVM & \small CV of SEISMIC \\
\hline
\small 200 &  \small 886 & \small 75.2\% & \small 75.0\% \\
\small 300 & \small 665 & \small 75.1\% & \small 79.0\% \\
\small 500 & \small 441 & \small 77.0\% & \small 77.2\% \\
\small 800 & \small 292 & \small 73.9\% & \small 78.0\% \\
\small 1000 & \small 241 & \small 73.1\% & \small 73.3\% \\
\end{tabular}
~\label{tab:svm}
\squeezeup
\end{table}

\subsection{Model Enhancement}
\par In this subsection, we identify key observations when applying SEISMIC to predict WeChat article final size and propose our approach to enhance the original SEISMIC model.

\par \textbf{Identifying Article Infectiousness} $p_t$. SEISMIC introduces a correction factor $\alpha$ on the estimated infectiousness to account for the information becoming stale and outdated. $\alpha$ decreases over time $t$ and scales down the estimated infectiousness in the future. This correction factor $\alpha$ is obtained via minimizing the average APE of a predefined trained tweet dataset. The variable $\alpha$ that decreases over time would inevitably lead to a ``supercritical'' state. However, in WeChat scenario, the infectiousness of articles varies because articles exhibit various propagation patterns. Therefore, the way to scale down the estimated infectiousness for future propagation should be strongly relevant to the property of the article itself. In other words, we should define a scaling factor to adjust infectiousness on the basis of the current propagation data of each article.
\par Many factors may affect article infectiousness, such as the article quality, the network structure, the current timestamp and the geographical locations of sharing users~\cite{SEISMIC:2015:KDD}. We leverage the propagation speed as
\begin{equation}
ps_{t} = (R_{t+1} - R_t)/R_f
\end{equation} to scale the estimated infectiousness, where $ps_{t}$ is the propagation speed at timestamp $t$, and $R_{t+1}$, $R_t$ and $R_f$ are the cumulative popularities at time $t+1$, $t$ and final size, respectively. Unlike SEISMIC that uses \textbf{an always decreasing $\alpha$} over time to scale down the infectiousness, we use propagation speed to adjust the original infectiousness at timestamp $t$, i.e.,
\begin{equation}
p_t' = (ps_t - ps_{min})/(ps_{max} - ps_{min}) \ast p_t
\end{equation} on the basis of the observations of some cascades of WeChat articles. In contrasts to tweets, articles are not necessarily stale over time; they may break out after some time.

\par \textbf{Estimating Mean Degree} $n_*$. In a point process-based model with constant infectiousness $p_t\equiv p$, a transition exists at a critical threshold $p^*$. Therefore, (1) If $p > p^*$, then $R_t \to \infty$ as $t \to \infty$ (This condition is called the \textit{supercritical} regime); (2) If $p < p^*$, then $sup_t$$R_t < \infty$ (This condition is called the \textit{subcritical} regime)~\cite{durrett:2010:probability}. SEISMIC defines the critical infectiousness threshold $p^*$ as $p^* = 1/n_*$, the reciprocal of the mean degree.
\par The collaboration with WeChat enables us to learn that the mean degree $n_*$ of WeChat network is 140. However, adopting this fixed value as $n_*$ for every article propagation prediction is far from adequate: \textbf{(1) Overlap of Users' Friends.} SEISMIC assumes that the friends of a user do not overlap with the friends of other users; however, when an article propagates over time, numerous users are actually exposed multiple times. \textbf{(2) Sharing Probability.} Some WeChat articles may cover a large scope of the underlying social network, whereas others may fail to penetrate a certain community of the network. Users may also have totally different tastes toward different types of articles, i.e., the sharing probability of WeChat users is different. Therefore, estimating an appropriate mean degree of the underlying social network in different propagation stages is indispensable for an accurate prediction. Besides, when the infectiousness $p_t$ is larger than $1/n_*$, the point process is ``supercritical'' and stays explosive, which does not exist in real-world propagation. Therefore, we adopt the previously speed-adjusted article infectiousness $p_t'$ to bound the searching space of an expected mean degree. With the bounded searching space $n_{*t} = 1/p_t'$, an appropriate mean degree can be estimated at different propagation stage.
\par Unlike the original SEISMIC, which only utilizes the instantaneous propagation information at the current timestamp and does not leverage the historical data that cover from the beginning to the current timestamp, our enhancement, i.e., using propagation speed to adjust article infectiousness and further bounding the searching space of the mean degree at each timestamp, fully leverages the information of ``what we have seen so far'' and makes the utmost of the historical propagation data. Fig.~\ref{fig:seismic} shows the first 24 h of the sharing activity of a popular article and the results of SEISMIC and our approach, WeSeer (introduced later). SEISMIC fails to give a prediction result until 2 h. At the $150^{th}$ min, the prediction is high and inaccurate (APE is about 3.5), although it later gives a good estimation of the final size. By contrast, WeSeer rapidly finds an accurate estimation of the final cascade size in only 20 min.

\subsection{Experts' Needs and Expectations}
\par We interview the experts (E.1-3) to identify their primary concerns and potential obstacles in their path regarding SEISMIC performance. At the end of the interviews, the need for a visual reasoning system to ground the team's conversation~\cite{li2018multi} with SEISMIC emerges as a key theme. Despite the differences in individual expectation for such a system, certain requirements are expressed across the board.
\par \textbf{R.1 Identifying effects of article infectiousness and mean network degree on final prediction results.} E.1-2 are interested in knowing how articles with different infectiousness propagate in the underlying social network and how the model responses and returns a prediction result. A clear visual cue of the best parameters of the predictive model should be displayed, i.e., the effects of the infectiousness and the mean network degree should be intuitively studied. A comparison of the outcomes should also be provided to understand the mechanism of the predictive model.
\par \textbf{R.2 Visualizing the propagation statistics and individual activities.} E.2 indicates that the statistical changes, e.g., pace and volume, of the propagation process of the articles throughout the observation period should be easily observed to identify the relationship between sharers' properties and the propagation prediction results. An intuitive and clear view should be provided to observe individual activities that may contribute to prediction outcomes for identifying their roles in article propagation process.
\par \textbf{R.3 Interactively Filtering Interesting Timeframes.} Historical sharing activities occur within different timeframes. Hence, interaction should be provided to select the interesting timeframes for a comparative investigation to further understand the factors that may affect prediction results.

\section{Workflow and Visual Design}
\begin{figure}[h]
    \centering
    \includegraphics[width=\linewidth]{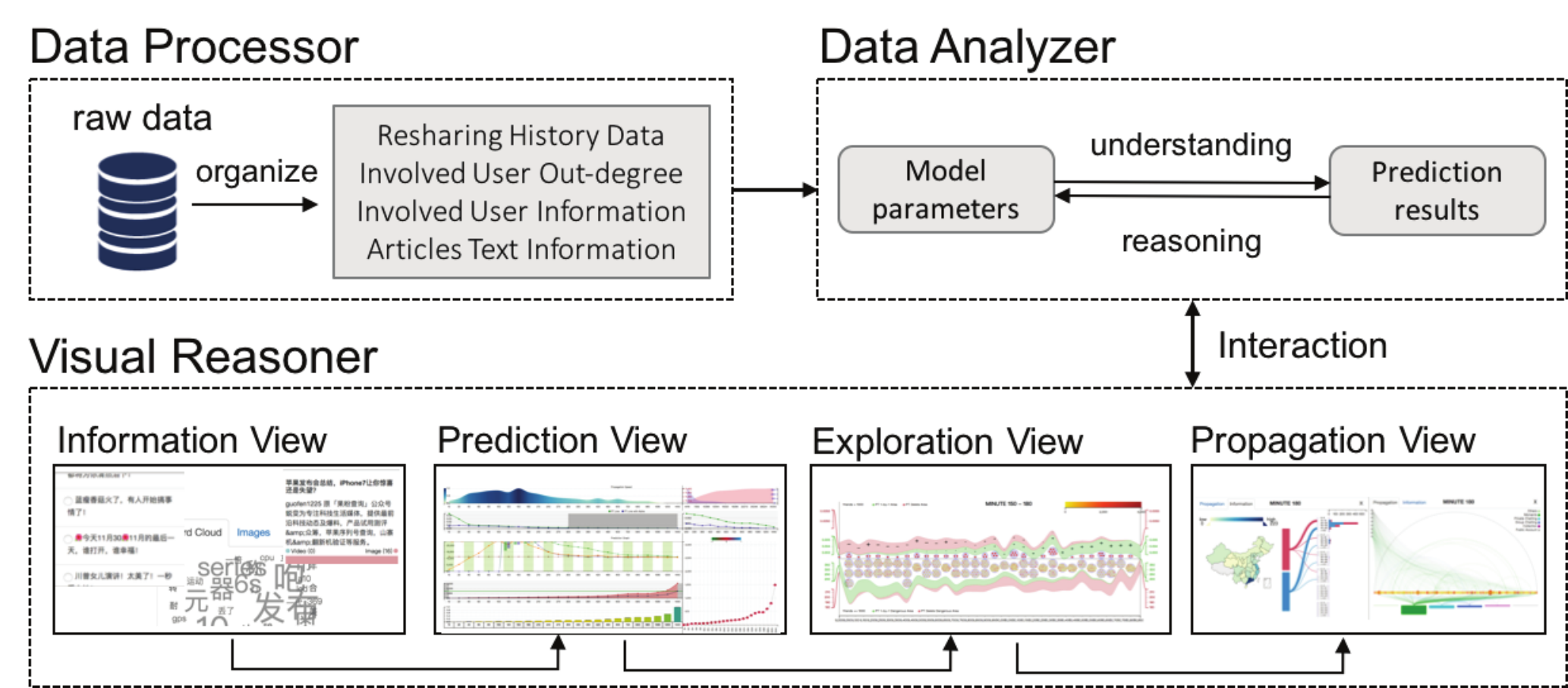} 
        \squeezeup
    \caption{Workflow of the proposed visual reasoning approach.}
    \label{fig:workflow}
    \squeezeup
\end{figure}

\begin{figure*}[h]
    \centering
    \includegraphics[width=\linewidth]{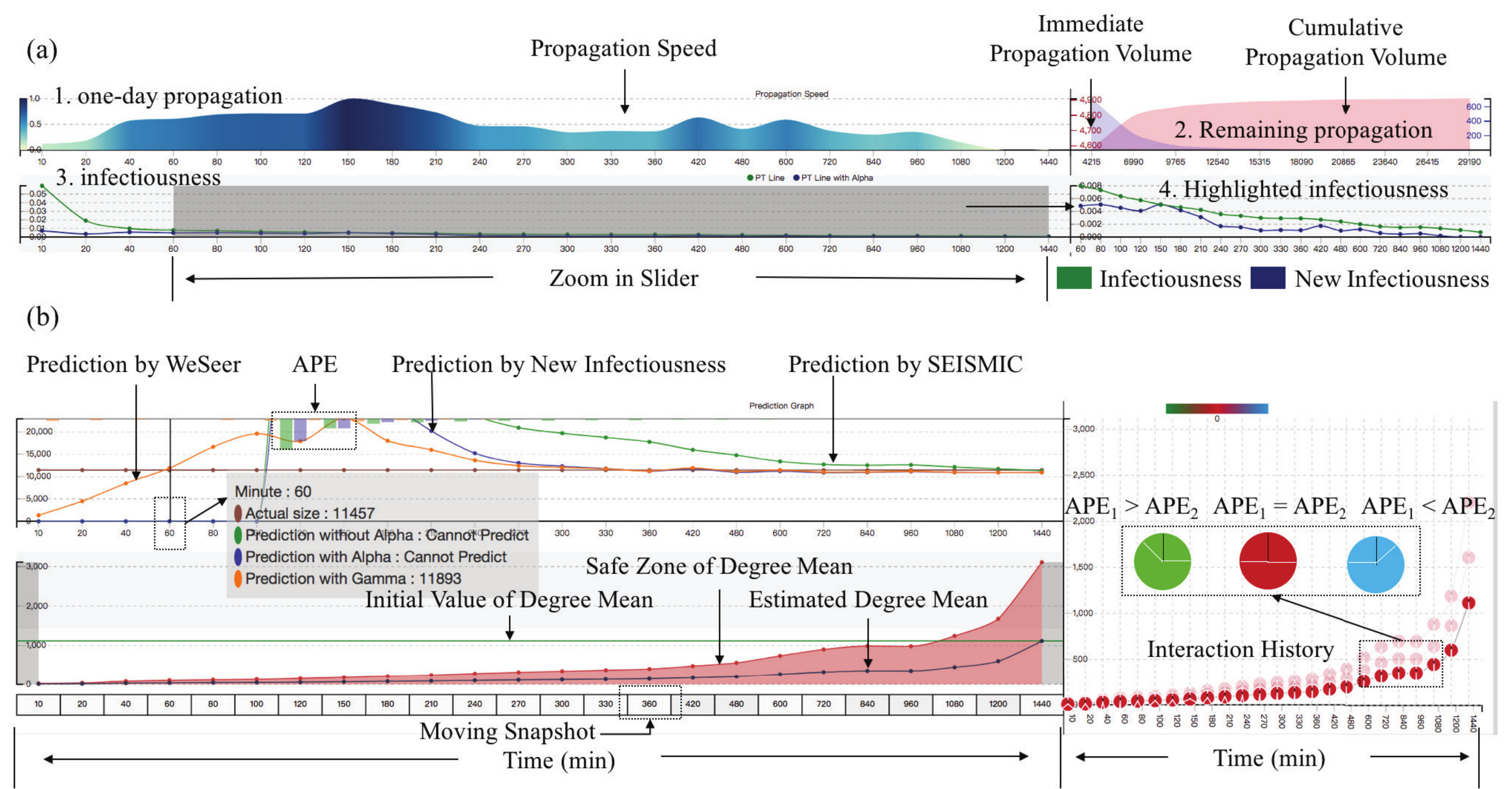} 
            \squeezeup
                   \squeezeup
    \caption{Prediction view: (a) Identifying the article infectiousness: (1) Propagation speed within a one-day time period. (2) Immediate propagation volume (purple area) and cumulative propagation volume (pink area) for the remaining propagation time period. (3) The original infectiousness (green line) and the new infectiousness (blue line) adjusted by the propagation speed. (4) Detailed comparison by zooming in the highlighted area in (3). (b) Users can initialize a value of mean degree and the system will automatically bound the values of mean degree at each timestamp. Three versions of prediction (i.e., by WeSeer, new infectiousness and SEISMIC) will be generated. User interactions will be reserved by faded curves.}
    \label{fig:predictionview}
    \squeezeup
\end{figure*}

\subsection{Workflow and Data Processing}
\par We design the workflow (Fig.~\ref{fig:workflow}) according to the experts' analysis process and partition it into three integral modules: the data processing module, the back-end enhanced model (i.e., the data analyzer module), and the front-end interactive interface (i.e., the visual reasoner module). In this section, we first introduce the data processing module, and we then present the visual reasoning system WeSeer in detail. The enhanced model can be referred to in Section 3.5.
\par In this paper, we use one-day historical propagation data to predict the final propagation size of articles. However, we can easily modify this setting in accordance with the scale of the historical propagation data we have. We unevenly divide the one-day time period into n consecutive timeframes ($t_1$ =0\~{}10 min, $t_2$ =10\~{}20 min, ..., $t_n$ =1200\~{}1440 min) based on the suggestions of the domain experts and then arrange the propagation data in the following aspects: \textbf{D.1 Sharing History:} We arrange the historical propagation data within each time window $t_i$ as a list, i.e., $res_{t_i}=[res_1, res_2, ...]$. The sharing data in each time window have the following format: $from_{uid}$ (parent user ID), $to_{uid}$ (current user ID), $from_{type}$ (parent propagation channel), $to_{type}$ (current sharing channel), and the current sharing timestamp; \textbf{D.2 Involved Users' Out-degree:} The out-degree $deg_u$ is defined as the number of friends of a user $u$. For all users, we have $deg_u$ = [$deg_{u_1}$, $deg_{u_2}$, ..., $deg_{u_n}$] in each timeframe; \textbf{D.3 Involved Users' Properties:} We extract the basic information of the users in each timeframe, such as gender, age, geographical information, and friend number. D.1-2 serve as the input for the predictive model (Prediction/Exploration View) and D.3 facilitates the exploration of the inherent information of the involved users (Propagation View).

\begin{figure*}[h]
    \centering
    \includegraphics[width=\linewidth]{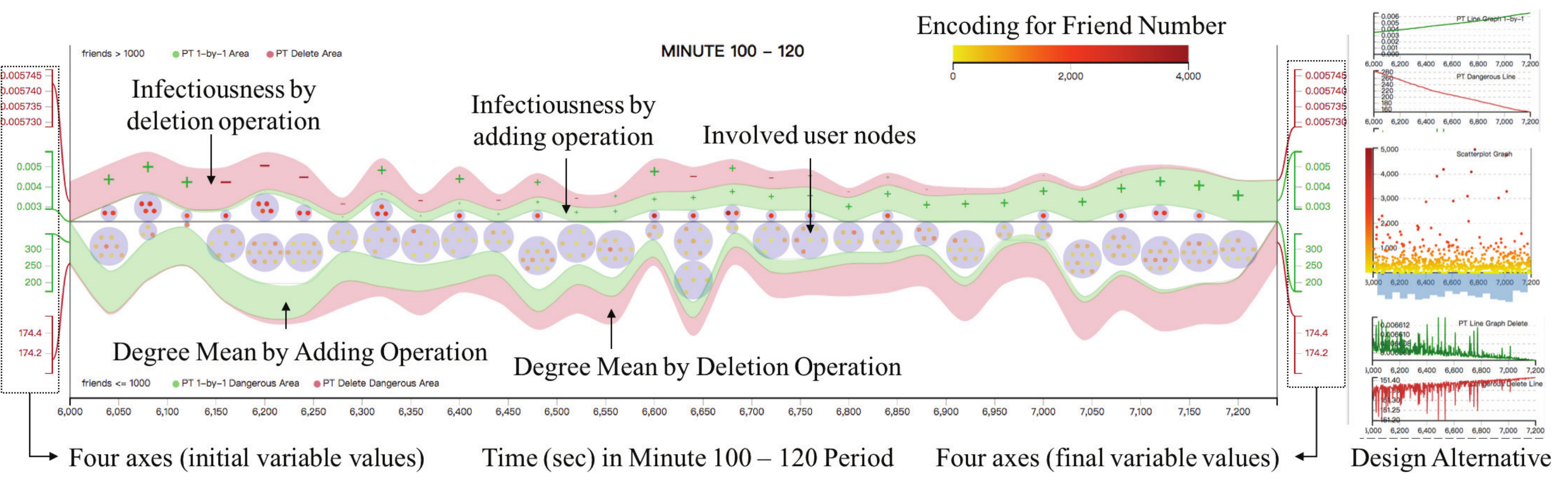} 
        \squeezeup
            \squeezeup
    \caption{Involved users are distributed along a horizontal axis that divides them into ``big nodes'' and ``small nodes''. Four variables, i.e., infectiousness and corresponding bounded mean degree derived from deletion and adding operations are rendered as river metaphors. We link them to four separate vertical axes, which represent initial values and final values of infectiousness by deletion, adding, mean degree by deletion, adding, respectively. `+' indicates an increasing infectiousness while `-' indicates a decreasing infectiousness. For example, in deletion operation, `+' means if without the corresponding nodes, infectiousness will increase; otherwise, it will decrease.}
    \label{fig:infectiousness}
    \squeezeup
\end{figure*}

\subsection{Visual Design}
\label{sec:design}
\par We develop a visual reasoning system, WeSeer, which consists of four views to help experts understand the model mechanism and the reasons for the propagation prediction result. The information view shows the basic information of the studied WeChat article and its publisher. The prediction view, the exploration view, and the propagation view help experts understand the underlying model and reason the prediction results. Once the team determines how visual reasoning can streamline their analysis, they can further apply their domain knowledge to improve prediction accuracy.

\subsubsection{Prediction View}
\par Prediction view provides comprehensive visual summary and understanding of the model mechanism, which visualizes the article infectiousness during the observation period and allows users to interactively fine-tune the model parameters in the bounded searching space of a reasonable mean degree. The target is to understand and compare the effects of article infectiousness and mean degree, thereby acquiring initial insights into the model.

\par \textbf{Visual Encoding for Speed-adjusted Article Infectiousness.} As discussed in Subsection 3.5, a reasonably varying infectiousness accounts for the phenomenon that the article is becoming stale and outdated. We use the propagation speed to adjust the article infectiousness (R.1, R.2). As shown in Fig.~\ref{fig:predictionview} (a) (1-2), we use deep blue to represent the highest propagation speed $ps_{max}$ in the observation period, and light blue represents the lowest propagation speed $ps_{min}$. More spaces are allocated to the one-day propagation data and less space to the immediate propagation out of the one-day data (represented by light purple) and the cumulative propagation over time (represented by pink color). We draw two infectiousness curves to represent the original $p_t$ and the one adjusted by propagation speed $p_t'$ (Fig.~\ref{fig:predictionview} (a) (3)). A slide is added to support detailed comparison on the highlighted infectiousness (Fig.~\ref{fig:predictionview} (a) (4)) due to the large-scale discrepancy between the two infectiousness curves.

\par \textbf{Visual Encoding for Reasoning Mean Degree.} The searching space of the mean degree should be displayed to reason an appropriate mean degree for the specific article (R.1). The suggested mean degree $n_*$ should always be bounded by $1/p_t'$ over time, which we render as the \textit{safe zone} (Fig.~\ref{fig:predictionview}) (b). Users can input a value as the initial mean network degree. With the boundary, the initial mean degree will be adjusted, thereby ensuring that the adjusted one at each timeframe is within the safe zone. We can then observe the distribution of APEs to determine the approximate mean degree at each timestamp. The users' exploration of the mean degree input will be preserved each time.

\par \textbf{Visual Encoding for Comparing Prediction Results.} We generate a new prediction curve (prediction by WeSeer) with the speed-adjusted article infectiousness and the estimated mean degree. Therefore, we have three versions of prediction results, i.e., the original SEISMIC, prediction by new infectiousness (adjusted by speed) and fixed mean degree (140), and WeSeer (prediction by new infectiousness and bounded mean degree). We encode the APE(s) at each timestamp of the three versions by three aligned bars (Fig.~\ref{fig:predictionview} (APE)) to compare them. We design a circular glyph at each timestamp to help users understand the potential propagation popularity by comparing $APE_1$ (comparing predictive size with one-day size) and $APE_2$ (comparing predictive size with the final size), as shown in Fig.~\ref{fig:predictionview}. The colors ranging from green to blue indicate the difference between the two APEs; if the difference approximates to zero, the color is red and the propagation of this article mainly occurs in the first day; otherwise, the color is green. Users can observe the distribution of APEs (the circular glyphs on all the faded curves) generated by multiple inputs of initial degree mean pre-setup by the system and determine the most appropriate mean degree at different stages of propagation.

\subsubsection{Exploration View}
\par We propose exploration view to clarify the contribution of individual users to the model performance (R.3), and support ``what-if'' analysis to help analysts understand the model. For a given timeframe, we obtain the collection of the sharing nodes within it and then conduct the following two operations: \textbf{(1) Deletion Operation.} For a particular timeframe, we delete one sharing record (an individual user) at a time, preserve the others, and calculate the corresponding infectiousness over time. \textbf{(2) Adding Operation.} Following a similar approach, we add the sharing node successively over time and examine the infectiousness change.
\par \textbf{Visual Encoding.} As shown in Fig.~\ref{fig:infectiousness}, we place the nodes within the corresponding timeframe in packed circles. We divide them by a horizontal axis. Nodes with more than 1,000 friends are placed in circles above the axis, and nodes with less than 1,000 friends are placed in circles below the axis. Nodes with 1,000 friends or above are defined as ``big nodes''; otherwise, they are defined as ``small nodes'' by the experts. We conduct a dynamic circle packing process, in which, if the space within the boundary is sufficient to generate a new packed circle, then we generate a new circle and evenly put the nodes into the circles; otherwise, we put all the nodes in one circle. Aside from all the involved nodes over time within the selected timeframe, this view also renders the changes in four variables, namely, the value of infectiousness from deletion ($p_{td}$) and adding ($p_{ta}$), and the value of bounded mean degree derived from deletion ($n_{*d}$) and adding ($n_{*a}$), as the width of four river metaphors. We link the rivers to four separate vertical axes to help analysts observe the initial and final variable values, i.e., $p_{td}$, $p_{ta}$, $n_{*d}$, and $n_{*a}$ for understanding how much each variable changes after this propagation timeframe. The sign of ``+'' indicates an increasing infectiousness, whereas ``-'' indicates a decreasing infectiousness. For example, in deletion operation, if we delete the corresponding nodes in the timeframe, then the infectiousness will either increase (``+''), or decrease (``-''). In adding operation, if we add the corresponding nodes in the timeframe, then the infectiousness will either increase (``+''), or decrease (``-'').

\par \textbf{Design Alternatives.} Originally, we use a scatter plot to show the distribution of involved users in the selected timeframe, with the $x$-axis representing the difference between the sharing timestamp and the post time of the article and the $y$-axis representing the number of friends. Four curves show the changes in the above four variables over the timeframe (Design Alternative in Fig.~\ref{fig:infectiousness}). However, we find that our analysts have to move their eyes, jumping from one view to another. ``\textit{Drawing the same x axis four times is duplicated for observation,}'' says E.2. The numerous nodes distributed in the scatter plot also cause visual clutter.

\begin{figure}[h]
    \centering
         \vspace{-2mm}
    \includegraphics[width=\linewidth]{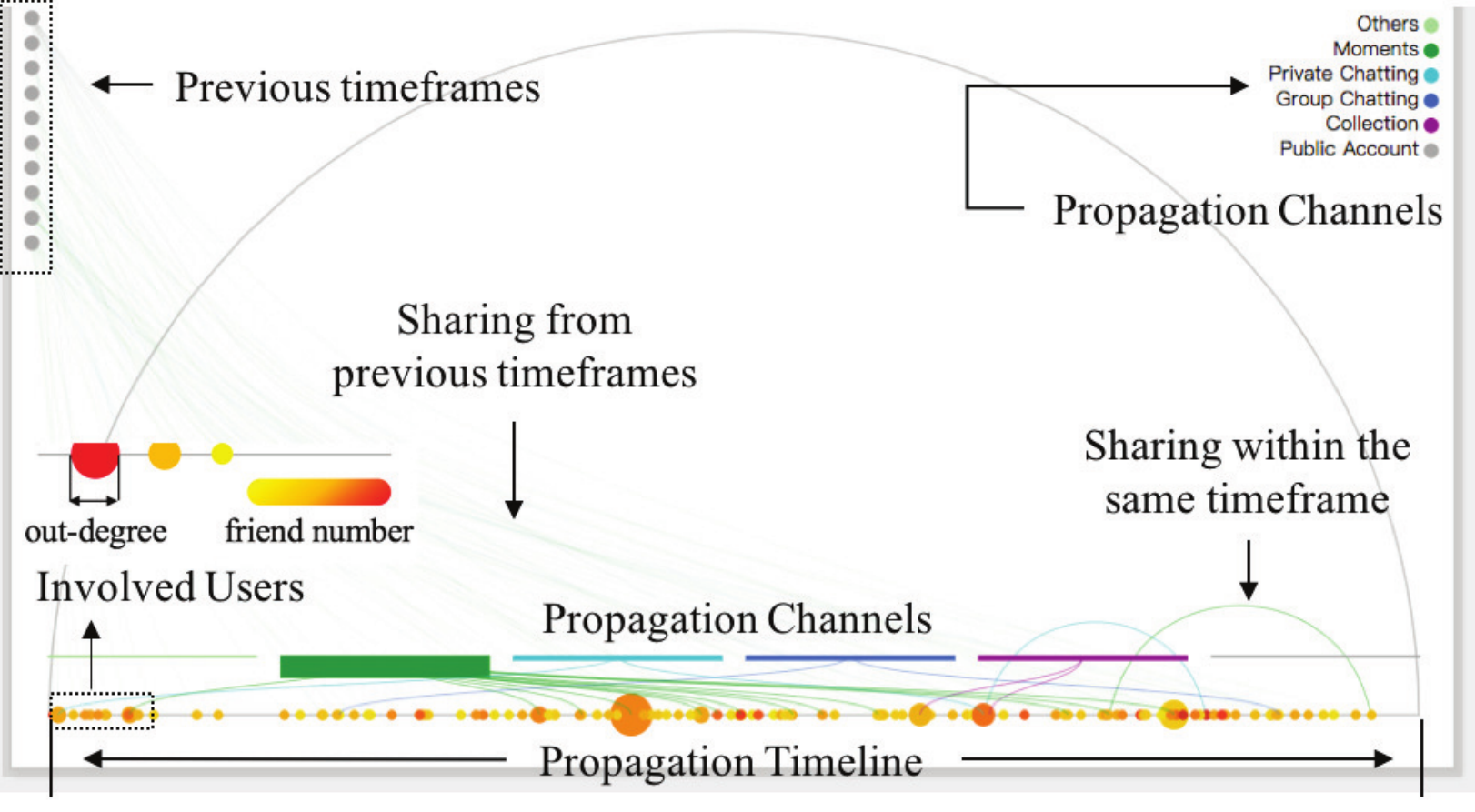} 
        \vspace{-6mm}
    \caption{Propagation view encodes the sharing processing using ``bible'' metaphor~\cite{kim2013case}. Sharers are represented by circles with size encoding degree and color encoding friend number. Sharing channels are shown with height encoding the number of shares that occur via the corresponding channel. Curve links from the parent to the child nodes. The parent and child nodes may locate in the same timeframe or the parent nodes may locate in the previous timeframe.}
    \label{fig:propagation}
  \vspace{-4mm}
\end{figure}

\begin{figure*}[h]
    \centering
    \includegraphics[width=\linewidth]{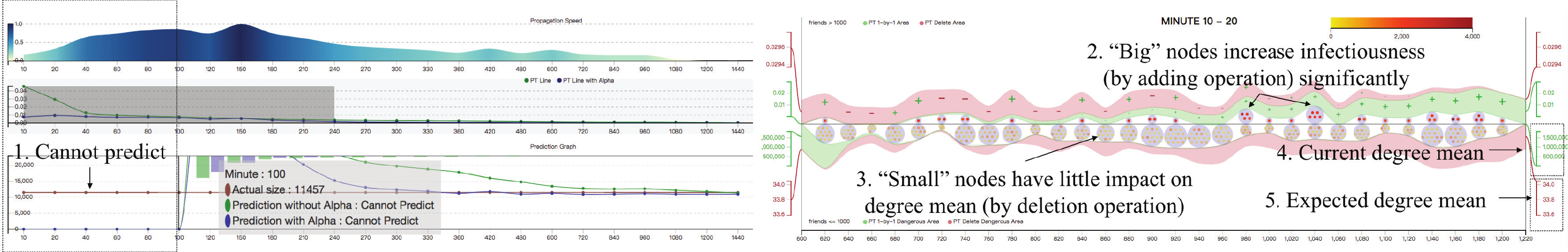} 
        \squeezeup
            \squeezeup
    \caption{(1) Both SEISMIC and the model adjusted only by speed fail to give a prediction result in the first two hours. (2) ``Big'' nodes increase infectiousness significantly. (3) ``Small nodes'' (see packed circles are mainly below 1000) and they have little impact on mean degree. (4) The current mean degree is too large and (5) the expected mean degree is around 22 after 10-20-minute time window of propagation.}
    \label{fig:caseone1}
    \squeezeup
\end{figure*}

\subsubsection{Propagation View}
\par Apart from the model, the inherent information of the involved users and their sharing status over the underlying network are also concerned by the experts. Therefore, we develop propagation view (Fig.~\ref{fig:propagation}) to show the relevant propagation and the involved users' information. This view demonstrates the distribution of the sharing channel and the sharing hierarchy. Some users in the current timeframe may share from the users in the previous timeframes; thereby, a previous timeframe indicator is added. Each vertically aligned gray node indicates one previous timeframe. In this manner, the communication among different timeframes can be easily observed and compared.

\subsection{Interactions Among the Views}
\par Our system provides rich interactions to facilitate efficient analysis: \textbf{(1) Linking and Brushing.} The system enables automatic linking among different views. For example, when users hover on a faded curve in prediction view, the corresponding setting of mean degree and prediction result will be displayed. When users want to inspect details of infectiousness in the prediction view, they can brush their interesting area of infectiousness. \textbf{(2) Details on Demand.} When users are interested in a particular timestamp, they can click the corresponding button, indicated as Fig.~\ref{fig:predictionview} (Moving Snapshot), and the detailed information, such as propagation users and user portrait, will be displayed. \textbf{(3) Hovering and Displaying.} Tooltips give users cues of the area of interest for facilitating further exploration. For example, in prediction view, concrete prediction results from SEISMIC and WeSeer are displayed as tooltips.

\section{Evaluation}
\par We first use several articles' propagation data to demonstrate the sensitivity and flexibility of WeSeer. We then automate our visual reasoning approach on a large collection of WeChat articles to quantitatively verify its efficacy. The showcased articles can be grouped into three categories, namely, \textit{celebrity} (introduction of some celebrities), \textit{event} (some emergency events), and ``\textit{chicken-soup}'' articles (e.g., festival greetings and inspirational articles). We also summarize the discussion and feedback from our experts.

\subsection{Case One: Highly Sensitive to Prediction}
\par In the first case, we show that WeSeer is more sensitive than SEISMIC (R.1). We leverage the propagation data of one ``event'' WeChat article, which introduces the release of iPhone 7. The final propagation size is 11500, and it lasts for more than three weeks.
\par \textbf{Prediction Performance Comparison.} As shown in Fig.~\ref{fig:caseone1} (1), we determine that in the first 2 h after the release of this article, the propagation speed is very high. SEISMIC and the model of infectiousness adjusted only by speed fail to give a prediction result. From the exploration view (Fig.~\ref{fig:caseone1} (right)), most sharing nodes are ``small'' nodes, and they exert limited impacts on the mean degree (by deletion operation) (the width of the pink band below the axis in Fig.~\ref{fig:caseone1} (3)). The number of ``big'' nodes that participated in the sharing process is insufficient; consequently, the infectiousness remains in a comparatively high-value scale (between 0.0294 and 0.0296). However, the ``big'' nodes can increase the article infectiousness (by adding operation) remarkably (the width of the green band above the axis) (Fig.~\ref{fig:caseone1} (2)). This view also gives us a visual cue that, after 10 min to 20 min time propagation, the current mean degree is over 500,000 (Fig.~\ref{fig:caseone1} (4)), but the expected mean degree is only approximately 33 (Fig.~\ref{fig:caseone1} (5)). However, in original SEISMIC, the mean degree value is fixed to 140, thereby failing to make a prediction.

\begin{figure}[h]
    \centering
    \includegraphics[width=\linewidth]{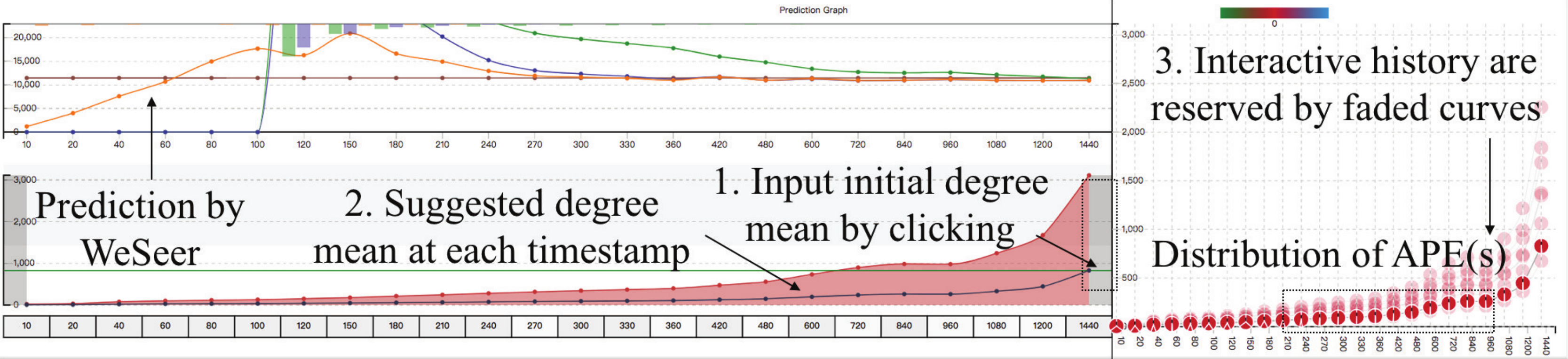} 
    \caption{(1) User randomly click and generate an initial value of mean degree, and the system suggests mean degree curve at each timestamp and gets the improved final prediction size (prediction by WeSeer). (3) The interactive selection of mean degree will be reserved and represented by faded curves. We can observe the distribution of APE(s) and know which timestamp returns the most accurate predictive result.}
    \label{fig:caseone2} 
\end{figure}

\par \textbf{Estimating Mean Degree.} WeSeer leverages the knowledge from the historical data (the one-day propagation data) to estimate an appropriate mean degree of the network for each timestamp. We click on the gray area (Fig.~\ref{fig:caseone2} (1)) and randomly select an initial value for the mean degree of the social network. WeSeer automatically suggests the mean degree $n_*$ curve at each timestamp (Fig.~\ref{fig:caseone2} (2)). Each interaction of users in selecting the initial value of $n_*$ is recorded for further comparison (Fig.~\ref{fig:caseone2} (3)). In this case, from the distribution of APE(s), we can observe that the appropriate mean degree recommended by WeSeer for this WeChat article is approximately 200, which enables us to attain a good prediction result at an early stage (Fig.~\ref{fig:caseone2} (distribution of APE(s))).

\begin{figure*}[h]
    \centering
    \includegraphics[width=\linewidth]{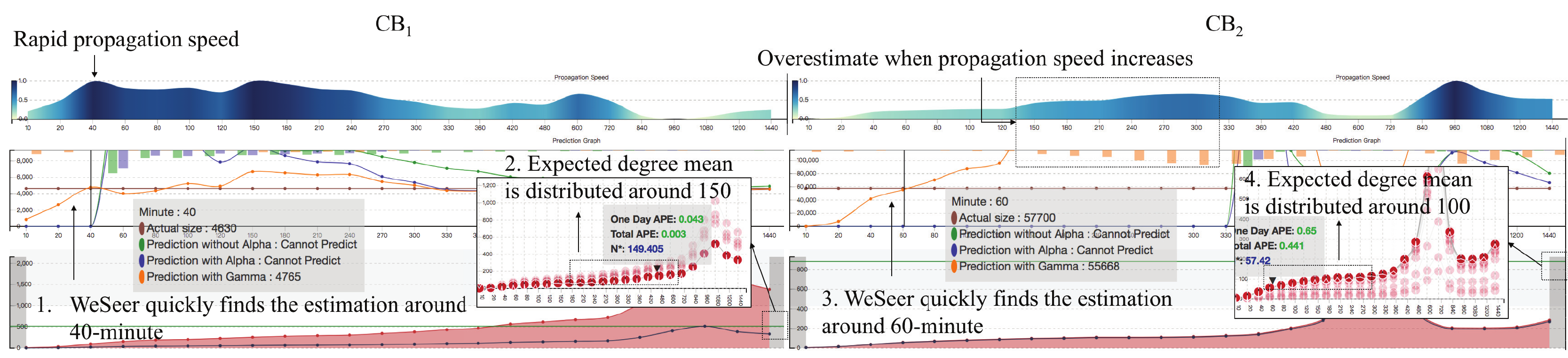} 
         \squeezeup
              \squeezeup
    \caption{(1) $CB_1$: WeSeer gives a prediction with a lower APE in 40-minute and (2) the recommendation mean degree for one-day propagation is around 150. (3) $CB_2$: WeSeer quickly finds the estimation around 60-minute and (4) the expected mean degree is distributed around 100. When the propagation speed increases, we may overestimate the final propagation size.}
    \label{fig:casetwo1}
     \squeezeup
\end{figure*}

\subsection{Case Two: Highly Flexible Toward Articles}
\par When E.1 is exploring the original model, he finds that some articles only need a few minutes of historical propagation data to estimate the final propagation popularity, whereas other articles consume more historical propagation data. For example, in the right part of Fig.~\ref{fig:seismic}, SEISMIC gives a prediction after 2h. He then expresses a desire to observe the prediction difference among articles. Therefore, in the second case, we leverage two \textit{celebrity} (CB) articles and one ``\textit{chicken-soup}'' (CS) WeChat article to demonstrate the flexibility of WeSeer toward different types of articles (R.1). 
\par We first give a brief introduction of the basic information of the three showcased articles: (1) $CB_1$ is about a celebrity, \textit{Jianlin Wang}, who is also a Chinese businessperson and the founder of Dalian Wanda Group. The content is that he was taking a TV interview and said that ``\textit{to achieve an ultimate life, we should first set up a small goal, for example, to earn a billion RMB first.}'' The propagation lasts for three days, and the original posted time is ``20160829 12:30:00.'' The final propagation size is 4630. (2) $CB_2$ is about a Chinese competitive female swimmer, \textit{Yuanhui Fu}, who specializes in the backstroke. At 2016 Summer Olympics in Rio, Fu gained popularity and became a swimming icon nationwide. Her series of facial expressions spread widely on the Internet, as well as her statement, ``\textit{I've already spent my supernatural energy.}'' The propagation lasts for three weeks. The original posted time is ``20160808 18:00:00,'' and the final propagation size is around 57700. (3) $CS$ is about greetings on the last day of November; e.g., ``\textit{Whoever reads this article on the last day of November will get blessed.}'' The original posted date is ``20161129,'' and the final propagation size is 110692.

\begin{figure}[h]
    \centering
    \includegraphics[width=\linewidth]{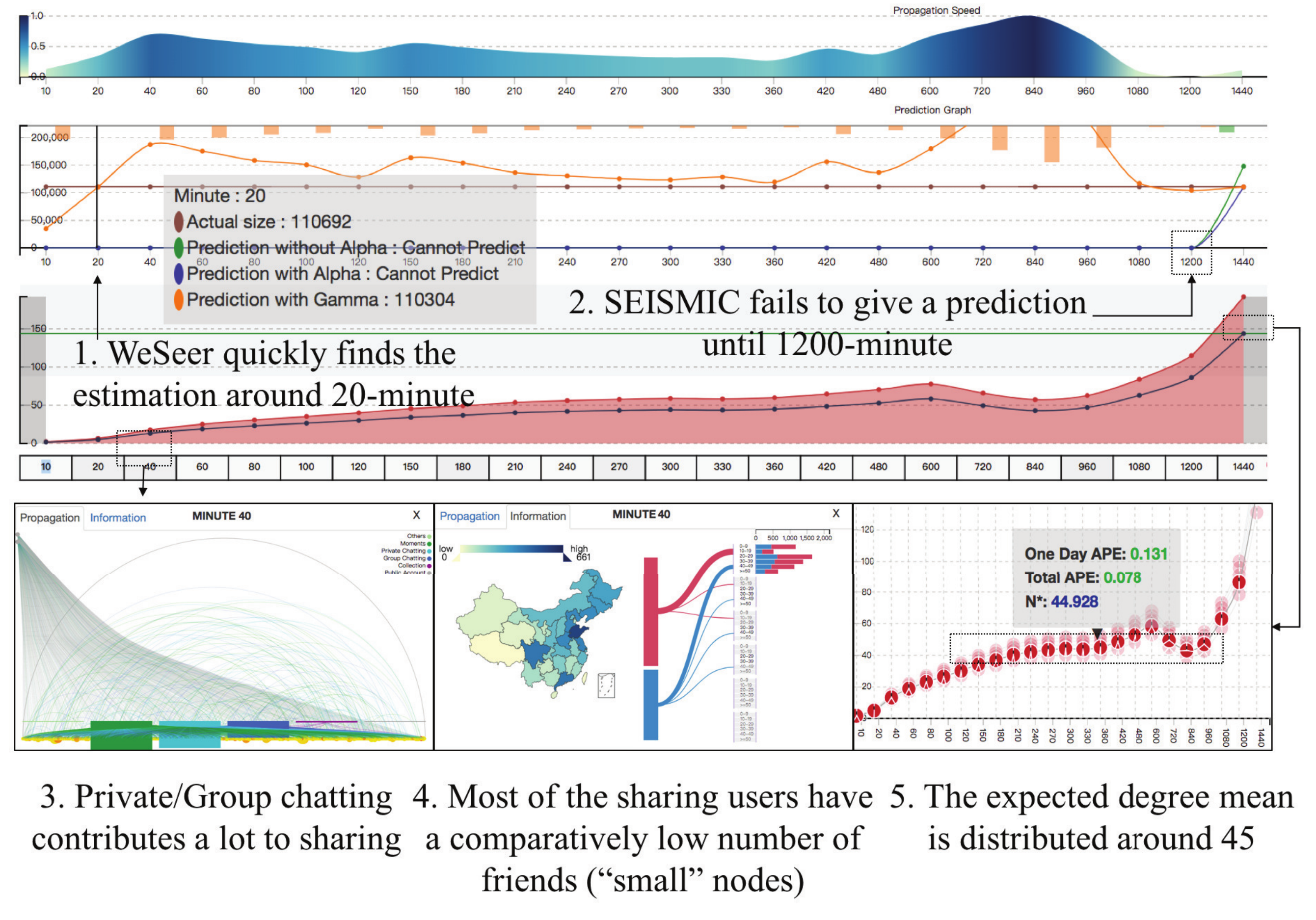} 
        \squeezeup
                \squeezeup
    \caption{(1) WeSeer immediately estimates a good result. (2) SEISMIC fails to give a prediction until the very end of one day. (3) Private/Group chatting occupy the $1^{st}$ place for propagation and (4) most of the users have a comparatively low number of friends. (5) The appropriate value for mean degree should be around 45.}
    \label{fig:casetwo2}
\end{figure}

\par \textbf{Prediction of Celebrity Articles.} As shown in Fig.~\ref{fig:casetwo1}, $CB_1$ exposes a rapid propagation speed in the initial stage. SEISMIC fails to give a predictive result until 60 min. WeSeer immediately finds the estimation accurately after approximately 40 min with a sufficiently small APE (Fig.~\ref{fig:casetwo1} (1)). The recommended mean degree is approximately 150 (Fig.~\ref{fig:casetwo1} (2)). $CB_2$ first presents a slow propagation speed but then begins to outbreak after several-hour propagation. We obtain the best predictive result with an APE of 0.441 after approximately 60 min (Fig.~\ref{fig:casetwo1} (3)), and the expected mean degree is approximately 100 (Fig.~\ref{fig:casetwo1} (4)). However, when the propagation speed increases, APE also increases and the predictive final propagation size is overestimated.

\par \textbf{Prediction of Chicken-soup Articles.} We observe different phenomena by analyzing a ``chicken-soup'' article (Fig.~\ref{fig:casetwo2}). The propagation lasts for more than a day but has a rapid propagation speed and a comparatively large coverage. This ``chicken-soup'' article has many sharing users who actively promote the propagation, given the propagation channel of Private Chatting ranks the first place (Fig.~\ref{fig:casetwo2} (3)). SEISMIC cannot give a prediction result until the very end of the one-day time period (Fig.~\ref{fig:casetwo2} (2)). Although the propagation covers the majority of mainland China and engages people in different ages, the average number of their friends is relatively low (0-1,000) (Fig.~\ref{fig:casetwo2} (4)). WeSeer can rapidly find an accurate estimation (Fig.~\ref{fig:casetwo2} (1)), and the majority of the adjusted mean degree across timeframes within the one-day period is only around 45 (Fig.~\ref{fig:casetwo2} (5)). E.1 comments that ``\textit{the quality of these articles is low, and they may therefore only propagate among certain communities of users and cannot spread to a larger group}''; ``\textit{Although the final size is over 100,000, unlike the `event' articles, the engaged users of `chicken-soup' articles are of homogeneity and lack diversification.}'' 

\subsection{Quantitative Evaluation}

\begin{figure}[h]
    \centering
    \includegraphics[width=\linewidth]{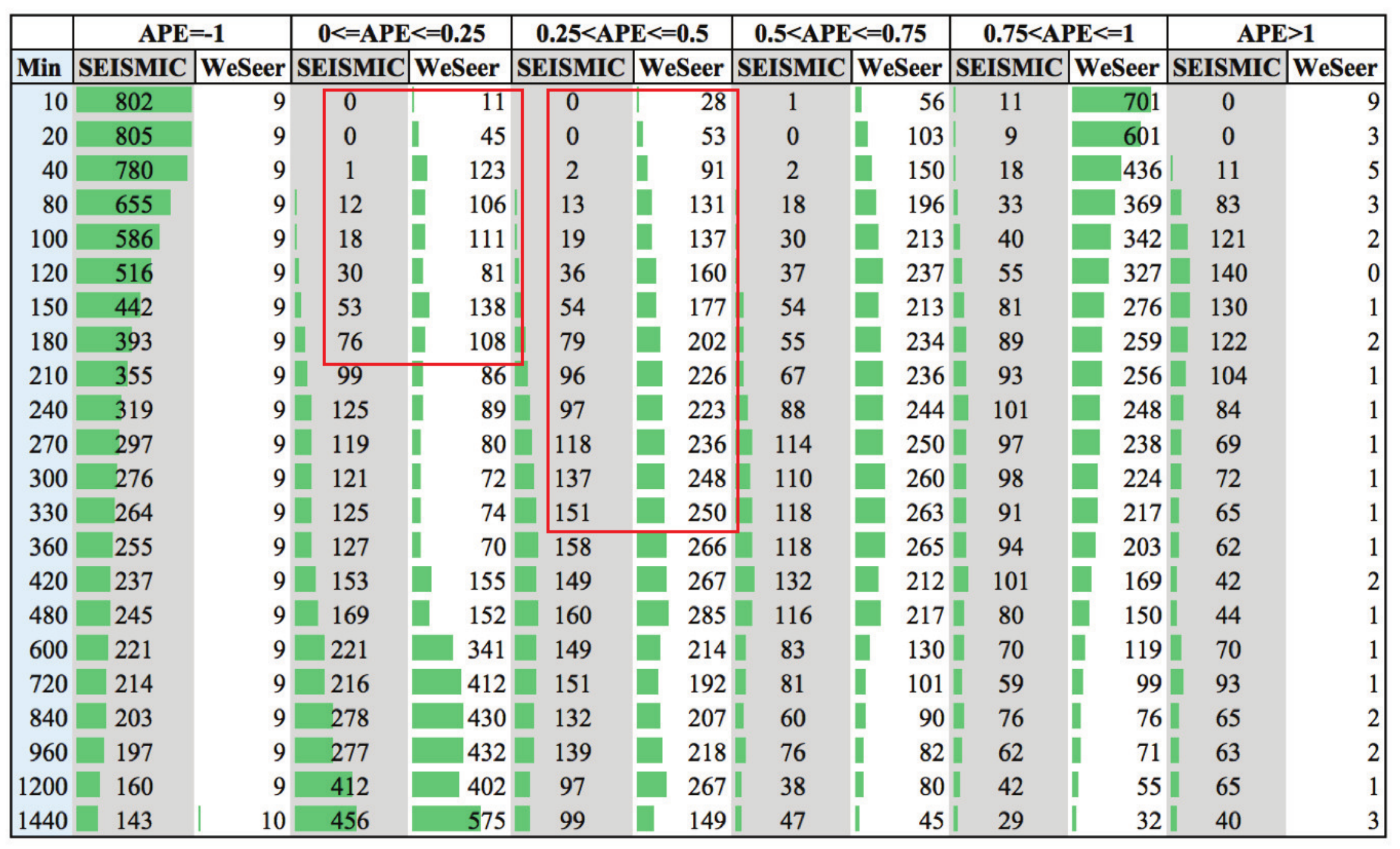} 
        \squeezeup

    \caption{Six categories of APEs and the corresponding number of predicted articles over time predicted by SEISMIC and WeSeer are plotted. X axis indicates the intervals of APE. Y axis indicates the time (min) used for prediction. The height of green bars inside shows the number of articles predicted by SEISMIC or WeSeer.}
    \label{fig:ape}
\end{figure}

\begin{figure*}[h]
    \centering
    \includegraphics[width=\linewidth]{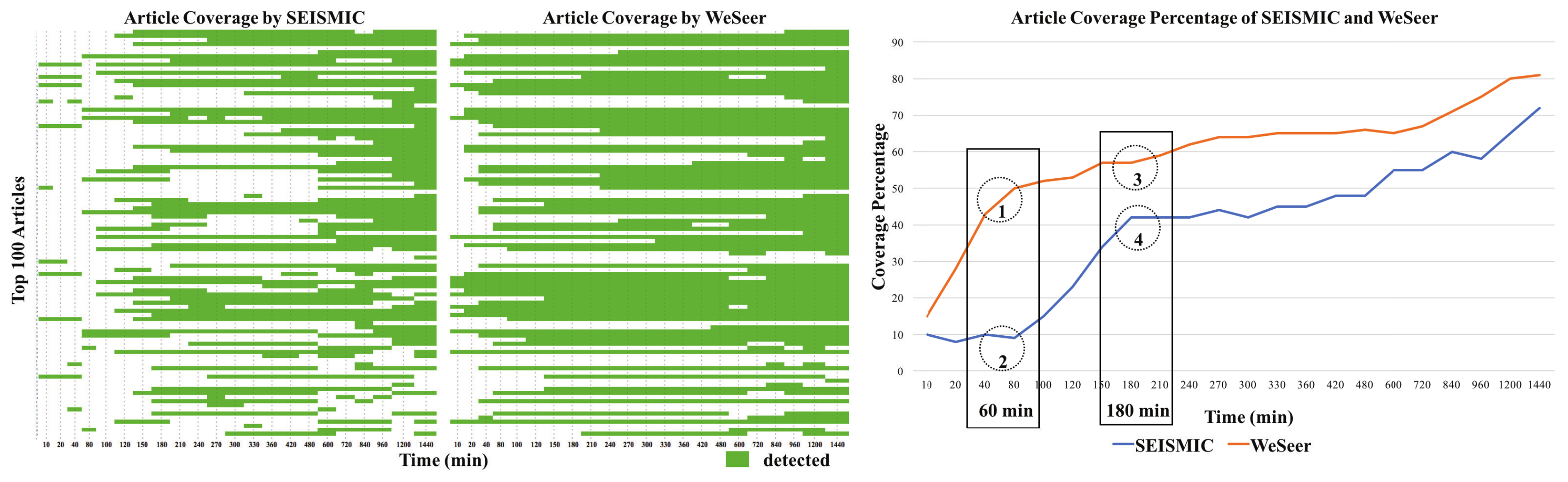} 
                \squeezeup
                            \squeezeup
    \caption{(Left): Coverage of top 100 most shared WeChat articles by SEISMIC and WeSeer. Each row represents an article and each column represents a time interval. White indicates that the predicted list of top 100 articles at time $t$ fails to cover a given article, and green indicates successful coverages. (Right): Comparison of the coverage percentage of top 100 articles between SEISMIC and WeSeer.}
    \label{fig:top100}
    \squeezeup
\end{figure*}

\par In this subsection, we quantify the performance of SEISMIC (original model) and WeSeer (enhanced model) in predicting WeChat article propagation on a large collection of WeChat articles. We randomly select 814 WeChat articles newly published on $21^{st}$, September 2017. The propagation of these articles all lasts for over a week. We first compare the prediction performance of SEISMIC and WeSeer via APE. We then identify whether WeSeer can perform better in identifying breakout articles than SEISMIC.

\par \textbf{Predicting Final Sharing Count.} We use the historical propagation data in 10, 20, 40, ..., 1440-min, respectively and run SEISMIC and WeSeer for each article. We obtain APE for each article and plot the distribution of APE(s) over time (10, 20, 40, ..., 1440-min) and the corresponding number of predicted articles that fall into that APE scale. As shown in Fig.~\ref{fig:ape}, SEISMIC often fails to give a prediction result (denoted by APE=-1) in each time interval. By contrast, WeSeer can give an estimation of the final propagation size (few APE=-1 occurs). Comparison of the trends of the number of predicted articles in each interval of APE between SEISMIC and WeSeer shows that WeSeer can predict more articles than SEISMIC, especially in the earlier stages. For example, the red rectangles highlight the WeSeer can cover more articles than SEISMIC at an earlier stage of prediction with APE falling into the scales of [0, 25] and (0.25, 0.5].

\par \textbf{Identifying Breakout Articles.} Identifying breakout WeChat articles is important in such tasks as trend forecasting or rumor detection. We rank a ground-truth list $L_M^*$, which contains top $M$ articles with the highest final sharing count. We generate the list of size $M$, $L_M^1(t)$ and $L_M^2(t)$, for SEISMIC and WeSeer, respectively. At each time $t$, the list contains the top $M$ articles with the highest predicted article counts at time $t$. We compare $L_M^1(t)$ and $L_M^2(t)$ with $L_M^*$ and calculate the \textit{BreakoutArticle Coverage}, which is defined by the proportion of articles in $L_M^*$ covered by $L_M^1(t)$ or $L_M^2(t)$. Fig.~\ref{fig:top100} compares the performance of SEISMIC and WeSeer in detecting the top 100 most shared articles and the corresponding coverage percentages over time. We can observe that WeSeer can predict and cover more articles that are in the real-world top 100 most shared list as there are more green blocks. From the right part of Fig.~\ref{fig:top100}, the circles highlight that around 50\% of articles (1) are covered in the first 60-min, overwhelming SEISMIC's only 10\% coverage (2). After 180 min, 60\% of the articles (3) can be covered by WeSeer, while for SEISMIC, only 40\% (4).

\subsection{Expert Review and Discussion}
\par We conduct a half an hour interview with our experts (E.1-3). They have been studying information propagation in WeChat for more than a year. We first demonstrate and explain our system, thereby allowing them to freely explore. We then have a post-interview with them. 

\par \textbf{System Capability.} All the experts appreciate the capability of WeSeer to support interactive exploration of propagation prediction results. E.1 comments that ``\textit{the smooth interactions make the exploration very effective and efficient.}'' They enjoy the prediction view that compares several versions of a predictive model simultaneously. ``\textit{WeSeer records the historical exploration results, thereby making the task of reasoning the mean degree considerably easy,}'' according to E.2.
\par \textbf{Learning Curve.} We involve the experts from the initial design of WeSeer. The main visualizations are based on their familiar visual metaphors, such as the curves and bars. After system briefing and explanations, the experts become familiar with the visualizations and the accompanied visual encodings. After they explore the prediction process of an article by themselves, they become thoroughly familiar with the interactions among the views. E.1-2 comment that ``\textit{the system helps us better understand the model and determine the possible effect of the parameters on the prediction result.}''
\par \textbf{System Outcome.} The experts are impressed by the outcome of WeSeer. Unlike the SEISMIC model that fails to sufficiently leverage the historical information to conduct prediction, WeSeer adjusts the article infectiousness by propagation speed and estimates mean degree via the adjusted infectiousness and can thus generate more sensitive and reliable prediction results. Therefore, WeSeer can cover articles that are considered in the ``supercritical state'' by the original SEISMIC model. They also appreciate that WeSeer combines the analysis of the prediction model with the propagation data to intuitively understand the panorama of propagation and the involved users, such as their portrait and sharing channels. E.2 comments that the quantitative comparison in the larger scale of articles further verifies the efficacy of WeSeer, and they all agree that our approach can automatically narrow down the searching space of the key parameters of the model. E.1-2 report that ``\textit{the low APE and the less time that the improved model needs to do prediction indicate that it works well.}'' E.3 states that ``\textit{the enhanced model can be practically used to replace the original one.}'' 
\par \textbf{Deployment and Factors behind Prediction.} The enhanced model has been deployed in WeChat for real-time predictive analysis and it generates a top k ranking list of articles with predictive propagation size in a descending order each day. The experts mainly select and inspect interesting articles from the daily ranking list. After exploring several articles, E.1-2 agree that involving ``big nodes'' in an earlier stage can significantly lower article infectiousness and can thus achieve a stable prediction result. The underlying network structure also greatly affects the propagation because the recommended mean degree for different types of articles is different. ``\textit{Articles have different capabilities to penetrate certain communities of the social network; for instance, `event' or `celebrity' articles have a larger mean degree than that of `chicken-soup' articles,}'' says E.1. E.2 comments that WeChat network is more private than other social networks; hence, Chatting plays important roles in certain types of articles' propagation, e.g., ``chicken-soup'' articles.

\subsection{Limitation}
\par Our work can be further improved through several aspects. \textbf{(1) Inductive article content.} Like SEISMIC that only utilizes the propagation data and basic properties of users, we do not utilize the content information of the articles, such as the title, the length of article content, and the article sentiment. \textbf{(2) Sharing probability.} WeSeer bounds the mean degree at each timestamp and concludes that articles differ in their capabilities to penetrate certain social communities. This conclusion is consistent with the observation that the users have their own preferences and have different propagation probabilities to share an article. If we can yield a sharing probability estimation from trained historical data, then we may have an improved prediction result. \textbf{(3) Effects of friend number and sharing channels.} If we can attain a complete WeChat social relationship, we can acquire a more accurate model by only computing the number of newly exposed users that each share brings in rather than directly using the degrees. \textbf{(4) Daily circle of posting/sharing.} Most WeChat users are located in China and Southeast Asia, unlike other social media users, such as Twitter and Facebook. A daily cycle of WeChat posting/sharing activities occurs, i.e., it may be reasonable that shares occur frequently at night and in the morning in the local time.

\par \textbf{``Slow Onset'' Articles.} When identifying breakout articles in the quantitative evaluation, the experts observe that some articles are not covered by SEISMIC and WeSeer in the prediction list of top 100 articles. For example, if the view count of an article surges two weeks after it was posted, it is less likely that we can predict the final propagation based merely on the propagation data from the first day. The experts comment that ``\textit{if an article's delayed outbreak happens long after the preset one-day propagation window, WeSeer may not be able to pick up any signal.}'' They further suggest that we can dynamically resize the historical propagation window to capture necessary information as the system input.

\par \textbf{Generality.} We also discuss with E.1-2 about the generality of our approach. They comment that the system, especially the parameter-relevant part, i.e., the prediction view, needs revision when applying to explore other point process-based models. Taking the aforementioned factors into consideration, such as the article content and sharing probability, may also generalize the current system.

\section{Conclusions and Future Work}
\par This study proposes WeSeer, a step-by-step interactive visual reasoning system, to help domain experts understand and improve a point process-based predictive model, SEISMIC, for a better prediction result. The reason for the factors behind WeChat articles' propagation prediction is also provided. Several showcased realistic WeChat articles and feedback from domain experts demonstrate the efficacy of our system. We also quantitatively verify the performance of WeSeer on a large scale of WeChat articles. In the future, we plan to improve the model and the system by considering the effects of article content, different sharing probabilities, newly exposing users and sharing channels, and the daily circle of posting/sharing activities. Another direction is that we can further design aggregated representations to identify certain subclasses of articles that propagate similarly (e.g., ``chicken-soup'' articles) and class-specific optimized parameters. Experts can then define appropriate training subsets and use them to derive fine-grained class-specific settings. Therefore, when making a prediction for an unknown article, the experts can find a matching class and apply appropriate optimized parameters accordingly. Furthermore, we currently mainly focus on two key components, namely, infectiousness and mean degree bounded at each timeframe during the observation period, and ignore other parameters, such as the memory kernels used for weighting the original infectiousness. We may further refine WeSeer by involving human expertise in analyzing the kernels to reach highly accurate predictions in the future.


%



 \section*{Acknowledgments}
 \par We thank the anonymous reviewers, WeChat experts
who participated in the studies, Dr. Yingcai WU for their valuable comments. This paper was supported by WeChat-HKUST Joint Lab on AI Technology (WHAT LAB) grant\#1617170-0.


\ifCLASSOPTIONcaptionsoff
  \newpage
\fi



%


\bibliographystyle{IEEEtran}
\bibliography{reference}

\begin{thebibliography}{10}
\providecommand{\url}[1]{#1}
\csname url@samestyle\endcsname
\providecommand{\newblock}{\relax}
\providecommand{\bibinfo}[2]{#2}
\providecommand{\BIBentrySTDinterwordspacing}{\spaceskip=0pt\relax}
\providecommand{\BIBentryALTinterwordstretchfactor}{4}
\providecommand{\BIBentryALTinterwordspacing}{\spaceskip=\fontdimen2\font plus
\BIBentryALTinterwordstretchfactor\fontdimen3\font minus
  \fontdimen4\font\relax}
\providecommand{\BIBforeignlanguage}[2]{{%
\expandafter\ifx\csname l@#1\endcsname\relax
\typeout{** WARNING: IEEEtran.bst: No hyphenation pattern has been}%
\typeout{** loaded for the language `#1'. Using the pattern for}%
\typeout{** the default language instead.}%
\else
\language=\csname l@#1\endcsname
\fi
#2}}
\providecommand{\BIBdecl}{\relax}
\BIBdecl

\bibitem{qiu2016lifecycle}
J.~Qiu, Y.~Li, J.~Tang, Z.~Lu, H.~Ye, B.~Chen, Q.~Yang, and J.~E. Hopcroft,
  ``The lifecycle and cascade of wechat social messaging groups,'' in
  \emph{Proceedings of the 25th International Conference on World Wide
  Web}.\hskip 1em plus 0.5em minus 0.4em\relax International World Wide Web
  Conferences Steering Committee, 2016, pp. 311--320.

\bibitem{Dove2017UX}
G.~Dove, K.~Halskov, J.~Forlizzi, and J.~Zimmerman, ``Ux design innovation:
  Challenges for working with machine learning as a design material,'' in
  \emph{CHI Conference on Human Factors in Computing Systems}, 2017, pp.
  278--288.

\bibitem{matsubara2012rise}
Y.~Matsubara, Y.~Sakurai, B.~A. Prakash, L.~Li, and C.~Faloutsos, ``Rise and
  fall patterns of information diffusion: model and implications,'' in
  \emph{Proceedings of the 18th ACM SIGKDD international conference on
  Knowledge discovery and data mining}.\hskip 1em plus 0.5em minus 0.4em\relax
  ACM, 2012, pp. 6--14.

\bibitem{mohler2011self}
G.~O. Mohler, M.~B. Short, P.~J. Brantingham, F.~P. Schoenberg, and G.~E. Tita,
  ``Self-exciting point process modeling of crime,'' \emph{Journal of the
  American Statistical Association}, vol. 106, no. 493, pp. 100--108, 2011.

\bibitem{SEISMIC:2015:KDD}
Q.~Zhao, M.~A. Erdogdu, H.~Y. He, A.~Rajaraman, and J.~Leskovec, ``Seismic: A
  self-exciting point process model for predicting tweet popularity,'' in
  \emph{Proceedings of the 21th ACM SIGKDD International Conference on
  Knowledge Discovery and Data Mining}.\hskip 1em plus 0.5em minus 0.4em\relax
  ACM, 2015, pp. 1513--1522.

\bibitem{gao2015modeling}
S.~Gao, J.~Ma, and Z.~Chen, ``Modeling and predicting retweeting dynamics on
  microblogging platforms,'' in \emph{Proceedings of the Eighth ACM
  International Conference on Web Search and Data Mining}.\hskip 1em plus 0.5em
  minus 0.4em\relax ACM, 2015, pp. 107--116.

\bibitem{ye:2016:visual}
P.~Ye, C.~Wang, Y.~Liu, Q.~Zhu, and K.~Zhang, ``Visual analysis of micro-blog
  retweeting using an information diffusion function,'' \emph{Journal of
  Visualization}, vol.~19, no.~4, pp. 823--838, 2016.

\bibitem{cao:2012:whisper}
N.~Cao, Y.-R. Lin, X.~Sun, D.~Lazer, S.~Liu, and H.~Qu, ``Whisper: Tracing the
  spatiotemporal process of information diffusion in real time,'' \emph{IEEE
  Transactions on Visualization and Computer Graphics}, vol.~18, no.~12, pp.
  2649--2658, 2012.

\bibitem{marcus:2011:twitinfo}
A.~Marcus, M.~S. Bernstein, O.~Badar, D.~R. Karger, S.~Madden, and R.~C.
  Miller, ``Twitinfo: aggregating and visualizing microblogs for event
  exploration,'' in \emph{Proceedings of the SIGCHI conference on Human factors
  in computing systems}.\hskip 1em plus 0.5em minus 0.4em\relax ACM, 2011, pp.
  227--236.

\bibitem{ho:2011:modeling}
C.-T. Ho, C.-T. Li, and S.-D. Lin, ``Modeling and visualizing information
  propagation in a micro-blogging platform,'' in \emph{Advances in Social
  Networks Analysis and Mining (ASONAM), 2011 International Conference
  on}.\hskip 1em plus 0.5em minus 0.4em\relax IEEE, 2011, pp. 328--335.

\bibitem{ozenc2010support}
F.~K. Ozenc, M.~Kim, J.~Zimmerman, S.~Oney, and B.~Myers, ``How to support
  designers in getting hold of the immaterial material of software,'' in
  \emph{Proceedings of the SIGCHI Conference on Human Factors in Computing
  Systems}.\hskip 1em plus 0.5em minus 0.4em\relax ACM, 2010, pp. 2513--2522.

\bibitem{cha:2009:measurement}
M.~Cha, A.~Mislove, and K.~P. Gummadi, ``A measurement-driven analysis of
  information propagation in the flickr social network,'' in \emph{Proceedings
  of the 18th international conference on World wide web}.\hskip 1em plus 0.5em
  minus 0.4em\relax ACM, 2009, pp. 721--730.

\bibitem{bao:2013:cumulative}
P.~Bao, H.-W. Shen, W.~Chen, and X.-Q. Cheng, ``Cumulative effect in
  information diffusion: empirical study on a microblogging network,''
  \emph{PloS one}, vol.~8, no.~10, p. e76027, 2013.

\bibitem{bao:2013:popularity}
P.~Bao, H.-W. Shen, J.~Huang, and X.-Q. Cheng, ``Popularity prediction in
  microblogging network: a case study on sina weibo,'' in \emph{Proceedings of
  the 22nd International Conference on World Wide Web}.\hskip 1em plus 0.5em
  minus 0.4em\relax ACM, 2013, pp. 177--178.

\bibitem{gao:2014:effective}
S.~Gao, J.~Ma, and Z.~Chen, ``Effective and effortless features for popularity
  prediction in microblogging network,'' in \emph{Proceedings of the 23rd
  International Conference on World Wide Web}.\hskip 1em plus 0.5em minus
  0.4em\relax ACM, 2014, pp. 269--270.

\bibitem{li:2016:exploring}
Q.~Li and Y.~Liu, ``Exploring the diversity of retweeting behavior patterns in
  chinese microblogging platform,'' \emph{Information Processing \&
  Management}, 2016.

\bibitem{tatar:2014:survey}
A.~Tatar, M.~D. de~Amorim, S.~Fdida, and P.~Antoniadis, ``A survey on
  predicting the popularity of web content,'' \emph{Journal of Internet
  Services and Applications}, vol.~5, no.~1, p.~8, 2014.

\bibitem{gao:2015:modeling}
S.~Gao, J.~Ma, and Z.~Chen, ``Modeling and predicting retweeting dynamics on
  microblogging platforms,'' in \emph{Proceedings of the Eighth ACM
  International Conference on Web Search and Data Mining}.\hskip 1em plus 0.5em
  minus 0.4em\relax ACM, 2015, pp. 107--116.

\bibitem{luo:2012:predicting}
Z.~Luo, Y.~Wang, and X.~Wu, ``Predicting retweeting behavior based on
  autoregressive moving average model,'' in \emph{International Conference on
  Web Information Systems Engineering}.\hskip 1em plus 0.5em minus 0.4em\relax
  Springer, 2012, pp. 777--782.

\bibitem{ding:2017:predicting}
X.~Ding and Y.~Tian, ``Predicting retweeting behavior based on bpnn in
  emergency incidents,'' \emph{Asia-Pacific Journal of Operational Research},
  vol.~34, no.~01, p. 1740011, 2017.

\bibitem{bao:2015:modeling}
P.~Bao, H.-W. Shen, X.~Jin, and X.-Q. Cheng, ``Modeling and predicting
  popularity dynamics of microblogs using self-excited hawkes processes,'' in
  \emph{Proceedings of the 24th International Conference on World Wide
  Web}.\hskip 1em plus 0.5em minus 0.4em\relax ACM, 2015, pp. 9--10.

\bibitem{bandari:2012:pulse}
R.~Bandari, S.~Asur, and B.~A. Huberman, ``The pulse of news in social media:
  Forecasting popularity,'' \emph{arXiv preprint arXiv:1202.0332}, 2012.

\bibitem{can:2013:predicting}
E.~F. Can, H.~Oktay, and R.~Manmatha, ``Predicting retweet count using visual
  cues,'' in \emph{Proceedings of the 22nd ACM international conference on
  Conference on information \& knowledge management}.\hskip 1em plus 0.5em
  minus 0.4em\relax ACM, 2013, pp. 1481--1484.

\bibitem{cheng2014can}
J.~Cheng, L.~Adamic, P.~A. Dow, J.~M. Kleinberg, and J.~Leskovec, ``Can
  cascades be predicted?'' in \emph{Proceedings of the 23rd international
  conference on World wide web}.\hskip 1em plus 0.5em minus 0.4em\relax ACM,
  2014, pp. 925--936.

\bibitem{naveed2011bad}
N.~Naveed, T.~Gottron, J.~Kunegis, and A.~C. Alhadi, ``Bad news travel fast: A
  content-based analysis of interestingness on twitter,'' in \emph{Proceedings
  of the 3rd International Web Science Conference}.\hskip 1em plus 0.5em minus
  0.4em\relax ACM, 2011, p.~8.

\bibitem{bakshy2011everyone}
E.~Bakshy, J.~M. Hofman, W.~A. Mason, and D.~J. Watts, ``Everyone's an
  influencer: quantifying influence on twitter,'' in \emph{Proceedings of the
  fourth ACM international conference on Web search and data mining}.\hskip 1em
  plus 0.5em minus 0.4em\relax ACM, 2011, pp. 65--74.

\bibitem{rodriguez2014uncovering}
M.~G. Rodriguez, J.~Leskovec, D.~Balduzzi, and B.~Sch{\"o}lkopf, ``Uncovering
  the structure and temporal dynamics of information propagation,''
  \emph{Network Science}, vol.~2, no.~1, pp. 26--65, 2014.

\bibitem{gomez2013structure}
M.~Gomez~Rodriguez, J.~Leskovec, and B.~Sch{\"o}lkopf, ``Structure and dynamics
  of information pathways in online media,'' in \emph{Proceedings of the sixth
  ACM international conference on Web search and data mining}.\hskip 1em plus
  0.5em minus 0.4em\relax ACM, 2013, pp. 23--32.

\bibitem{keim:2015:bridging}
D.~A. Keim, T.~Munzner, F.~Rossi, and M.~Verleysen, ``Bridging information
  visualization with machine learning (dagstuhl seminar 15101),'' in
  \emph{Dagstuhl Reports}, vol.~5, no.~3.\hskip 1em plus 0.5em minus
  0.4em\relax Schloss Dagstuhl-Leibniz-Zentrum fuer Informatik, 2015.

\bibitem{li2013visual}
Q.~Li, H.~Qu, L.~Chen, R.~Wang, J.~Yong, and D.~Si, ``Visual analysis of
  retweeting propagation network in a microblogging platform,'' in
  \emph{Proceedings of the 6th international symposium on visual information
  communication and interaction}.\hskip 1em plus 0.5em minus 0.4em\relax ACM,
  2013, pp. 44--53.

\bibitem{chen2016d}
S.~Chen, S.~Chen, Z.~Wang, J.~Liang, X.~Yuan, N.~Cao, and Y.~Wu, ``D-map:
  Visual analysis of ego-centric information diffusion patterns in social
  media,'' in \emph{Visual Analytics Science and Technology (VAST), 2016 IEEE
  Conference on}.\hskip 1em plus 0.5em minus 0.4em\relax IEEE, 2016, pp.
  41--50.

\bibitem{lu2016exploring}
Y.~Lu, M.~Steptoe, S.~Burke, H.~Wang, J.-Y. Tsai, H.~Davulcu, D.~Montgomery,
  S.~R. Corman, and R.~Maciejewski, ``Exploring evolving media discourse
  through event cueing,'' \emph{IEEE transactions on visualization and computer
  graphics}, vol.~22, no.~1, pp. 220--229, 2016.

\bibitem{zhao2014fluxflow}
J.~Zhao, N.~Cao, Z.~Wen, Y.~Song, Y.-R. Lin, and C.~Collins, ``\# fluxflow:
  Visual analysis of anomalous information spreading on social media,''
  \emph{IEEE Transactions on Visualization and Computer Graphics}, vol.~20,
  no.~12, pp. 1773--1782, 2014.

\bibitem{wu2014opinionflow}
Y.~Wu, S.~Liu, K.~Yan, M.~Liu, and F.~Wu, ``Opinionflow: Visual analysis of
  opinion diffusion on social media,'' \emph{IEEE Transactions on Visualization
  and Computer Graphics}, vol.~20, no.~12, pp. 1763--1772, 2014.

\bibitem{chen2017social}
S.~Chen, L.~Lin, and X.~Yuan, ``Social media visual analytics,'' in
  \emph{Computer Graphics Forum}, vol.~36, no.~3.\hskip 1em plus 0.5em minus
  0.4em\relax Wiley Online Library, 2017, pp. 563--587.

\bibitem{el:2014:predictive}
M.~El-Assady, W.~Jentner, M.~Stein, F.~Fischer, T.~Schreck, and D.~Keim,
  ``Predictive visual analytics: Approaches for movie ratings and discussion of
  open research challenges,'' in \emph{An IEEE VIS 2014 Workshop: Visualization
  for Predictive Analytics}, 2014.

\bibitem{gleicher:2014:position}
M.~Gleicher, ``Position paper: Towards comprehensible predictive modeling,'' in
  \emph{Proceedings of IEEE VIS Workshop: Visualization for Predictive
  Analytics}, 2014.

\bibitem{muhlbacher2014opening}
T.~M{\"u}hlbacher, H.~Piringer, S.~Gratzl, M.~Sedlmair, and M.~Streit,
  ``Opening the black box: Strategies for increased user involvement in
  existing algorithm implementations,'' \emph{IEEE transactions on
  visualization and computer graphics}, vol.~20, no.~12, pp. 1643--1652, 2014.

\bibitem{yeon2015predictive}
H.~Yeon and Y.~Jang, ``Predictive visual analytics using topic composition,''
  in \emph{Proceedings of the 8th international symposium on visual information
  communication and interaction}.\hskip 1em plus 0.5em minus 0.4em\relax ACM,
  2015, pp. 1--8.

\bibitem{maciejewski2011forecasting}
R.~Maciejewski, R.~Hafen, S.~Rudolph, S.~G. Larew, M.~A. Mitchell, W.~S.
  Cleveland, and D.~S. Ebert, ``Forecasting hotspots—a predictive analytics
  approach,'' \emph{IEEE Transactions on Visualization and Computer Graphics},
  vol.~17, no.~4, pp. 440--453, 2011.

\bibitem{hao:2011:visual}
M.~C. Hao, H.~Janetzko, S.~Mittelst{\"a}dt, W.~Hill, U.~Dayal, D.~A. Keim,
  M.~Marwah, and R.~K. Sharma, ``A visual analytics approach for
  peak-preserving prediction of large seasonal time series,'' in \emph{Computer
  Graphics Forum}, vol.~30, no.~3.\hskip 1em plus 0.5em minus 0.4em\relax Wiley
  Online Library, 2011, pp. 691--700.

\bibitem{bosch2013scatterblogs2}
H.~Bosch, D.~Thom, F.~Heimerl, E.~P{\"u}ttmann, S.~Koch, R.~Kr{\"u}ger,
  M.~W{\"o}rner, and T.~Ertl, ``Scatterblogs2: Real-time monitoring of
  microblog messages through user-guided filtering,'' \emph{IEEE Transactions
  on Visualization and Computer Graphics}, vol.~19, no.~12, pp. 2022--2031,
  2013.

\bibitem{malik2014proactive}
A.~Malik, R.~Maciejewski, S.~Towers, S.~McCullough, and D.~S. Ebert,
  ``Proactive spatiotemporal resource allocation and predictive visual
  analytics for community policing and law enforcement,'' \emph{IEEE
  transactions on visualization and computer graphics}, vol.~20, no.~12, pp.
  1863--1872, 2014.

\bibitem{Lu2017The}
Y.~Lu, R.~Garcia, B.~Hansen, M.~Gleicher, and R.~Maciejewski, ``The
  state-of-the-art in predictive visual analytics,'' in \emph{Computer Graphics
  Forum}, 2017, pp. 539--562.

\bibitem{muhlbacher2013partition}
T.~M{\"u}hlbacher and H.~Piringer, ``A partition-based framework for building
  and validating regression models,'' \emph{IEEE Transactions on Visualization
  and Computer Graphics}, vol.~19, no.~12, pp. 1962--1971, 2013.

\bibitem{durrett:2010:probability}
R.~Durrett, \emph{Probability: theory and examples}.\hskip 1em plus 0.5em minus
  0.4em\relax Cambridge university press, 2010.

\bibitem{crane:2008:robust}
R.~Crane and D.~Sornette, ``Robust dynamic classes revealed by measuring the
  response function of a social system,'' \emph{Proceedings of the National
  Academy of Sciences}, vol. 105, no.~41, pp. 15\,649--15\,653, 2008.

\bibitem{li2018multi}
Q.~Li, Z.~Wu, P.~Xu, H.~Qu, and X.~Ma, ``A multi-phased co-design of an
  interactive analytics system for moba game occurrences,'' in
  \emph{Proceedings of the 2018 on Designing Interactive Systems Conference
  2018}.\hskip 1em plus 0.5em minus 0.4em\relax ACM, 2018, pp. 1321--1332.

\bibitem{kim2013case}
H.~Kim and J.~W. Park, ``Case analysis of bible visualization based on text
  data traits-focused on content, structure, quotation of text,'' \emph{The
  Journal of the Korea Contents Association}, vol.~13, no.~8, pp. 83--92, 2013.

\end{thebibliography}

%

\begin{IEEEbiography}[{\includegraphics[width=1in,height=1.25in,clip,keepaspectratio]{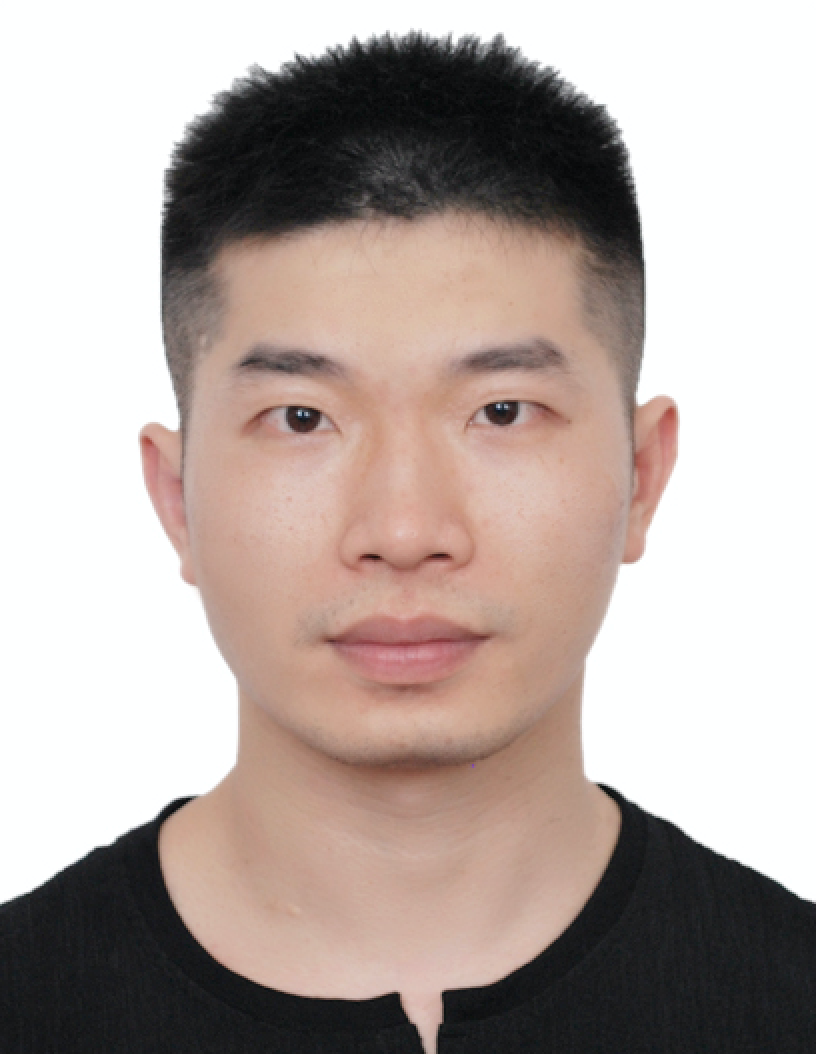}}]{Quan Li}
\par is a PhD candidate at the Hong Kong University of Science and Technology. His research interest includes visualization and human-computer interaction, with focuses on virtual environment visualization like gameplay visual analytics, social media analysis, and explainable machine learning. He obtained his Bachelor's Degree from Wuhan University in 2009 and Master's Degree from Tsinghua University.
\end{IEEEbiography}

\vspace{-12mm}

\begin{IEEEbiography}[{\includegraphics[width=1in,height=1.25in,clip,keepaspectratio]{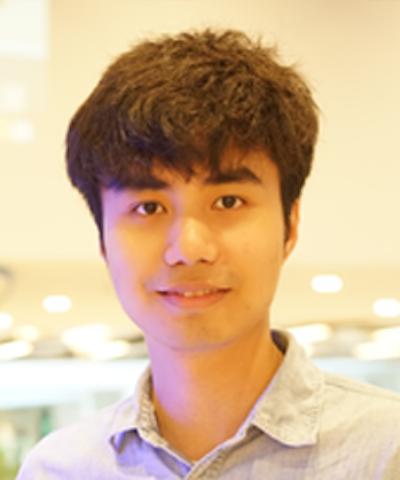}}]{Ziming Wu}
\par is a PhD candidate at the Hong Kong University of Science and Technology. His research interest includes human computer interaction and data-driven design. He received his Bachelor's Degree in Computer Science from South China University of Technology.
\end{IEEEbiography}

\vspace{-12mm}
\begin{IEEEbiography}[{\includegraphics[width=1in,height=1.25in,clip,keepaspectratio]{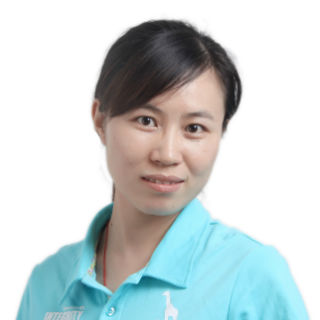}}]{Lingling Yi}
\par is a senior researcher at Tencent. She obtained her Master's Degree  in mathematics department from South China University of Technology in 2009. After joining Tencent, she has been engaged in data mining relevant work. She is currently responsible for WeChat social work such as ``lookalike'' and information dissemination. She has led several projects in Tencent such as APP social recommendation, friend recommendation, user social circle mining, and user portrait construction. 
\end{IEEEbiography}

\vspace{-12mm}

\begin{IEEEbiography}[{\includegraphics[width=1in,height=1.25in,clip,keepaspectratio]{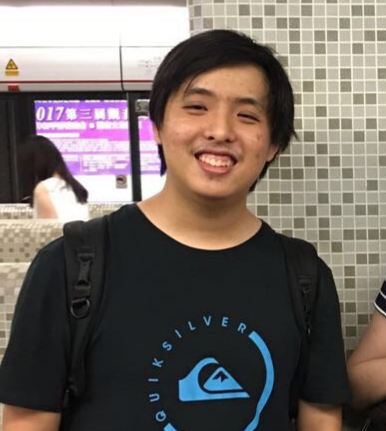}}]{Kristanto Sean Njotoprawiro}
\par is a research assistant at the Hong Kong University of Science and Technology. He received a Bachelor's Degree in Information Technology from Ciputra University of Indonesia in August 2013. He also received a Master's Degree of Science in Information Technology from the Hong Kong University of Science and Technology in August 2016.
\end{IEEEbiography}

\vspace{-12mm}

\begin{IEEEbiography}[{\includegraphics[width=1in,height=1.25in,clip,keepaspectratio]{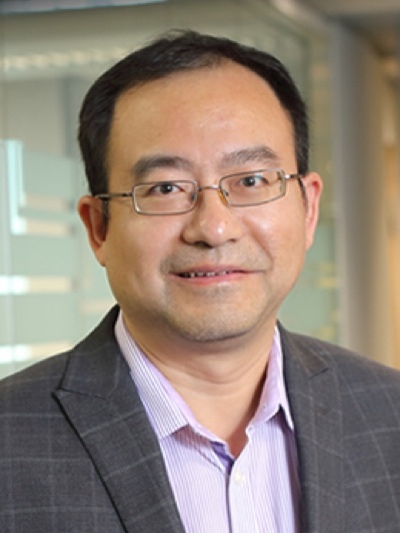}}]{Huamin Qu}
\par is a full professor in the Department of Computer Science and Engineering at the Hong Kong University of Science and Technology. His main research interests are in visualization and human-computer interaction, with focuses on urban informatics, social network analysis, e-learning, text visualization, and explainable artificial intelligence. He obtained BS in Mathematics from Xi'an Jiaotong University, China, MS and a PhD in Computer Science from the Stony Brook University.
\end{IEEEbiography}

\vspace{-12mm}
\begin{IEEEbiography}[{\includegraphics[width=1in,height=1.25in,clip,keepaspectratio]{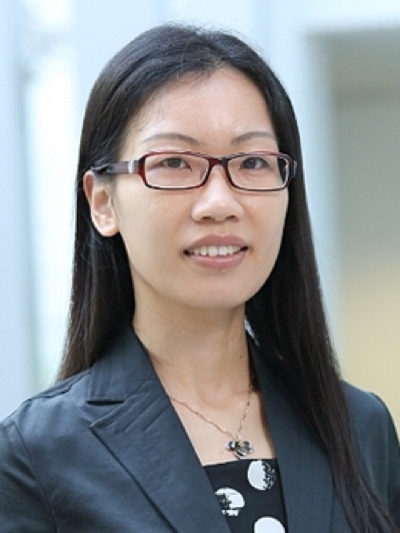}}]{Xiaojuan Ma}
\par is an assistant professor of Human-Computer Interaction (HCI) at the Department of Computer Science and Engineering, Hong Kong University of Science and Technology. She received her PhD in Computer Science at Princeton University. She was a post-doctoral researcher at the Human-Computer Interaction Institute of Carnegie Mellon University, and before that a research fellow in the National University of Singapore in the Information Systems department. Before joining HKUST, she was a researcher of Human-Computer Interaction at Noah's Ark Lab, Huawei Tech. Investment Co., Ltd. in Hong Kong.
\end{IEEEbiography}




\end{document}